\documentclass[twocolumn,epjc3]{svjour3}
\usepackage[dvipsnames]{xcolor}
\usepackage{booktabs, cite, feynmp-auto, graphicx, hyperref, listings, newtxtext, newtxmath, nicefrac, relsize, xspace}
\hypersetup{colorlinks=true, linkcolor=BurntOrange, citecolor=BurntOrange, filecolor=BurntOrange, urlcolor=BurntOrange}
\journalname{Eur. Phys. J. C}
\newcommand{\orcid}[1]{\begingroup\href{https://orcid.org/#1}{\includegraphics[width=9pt]{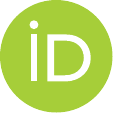}} \endgroup}
\newcommand{\be}{\begin{equation}}
\newcommand{\ee}{\end{equation}}
\def\bpm{\begin{pmatrix}}
\def\epm{\end{pmatrix}}
\newcommand{\ie}{\textit{i.e.}}
\def\etc{\textit{etc.}}
\def\eg{\textit{e.g.}}
\newcommand{\mgamc}{{\sc MG5\_aMC}\xspace}
\newcommand{\py}{{\sc Pythia~8}\xspace}
\def\madanalysis{{\sc MadAnalysis~5}\xspace}
\def\fastjet{{\sc FastJet}\xspace}
\DeclareUnicodeCharacter{2212}{-}

\begin{document}
\title{Simulating toponium formation signals at the LHC}

\author{
  Benjamin Fuks\,\orcid{0000-0002-0041-0566}\thanksref{e1,add1},
  Kaoru Hagiwara\thanksref{e2,add2},
  Kai Ma\,\orcid{0000-0001-7119-6117}\thanksref{e3,add5}
  and Ya-Juan Zheng\,\orcid{0000-0003-2961-5378}\thanksref{e4,add6}
}

\thankstext{e1}{\href{mailto:fuks@lpthe.jussieu.fr}{fuks@lpthe.jussieu.fr}}
\thankstext{e2}{\href{mailto:kaoru.hagiwara@kek.jp}{kaoru.hagiwara@kek.jp}}
\thankstext{e3}{\href{mailto:makai@ucas.ac.cn}{makai@ucas.ac.cn}}
\thankstext{e4}{\href{mailto:yjzheng@iwate-u.ac.jp}{yjzheng@iwate-u.ac.jp}}

\institute{
  Laboratoire de Physique Théorique et Hautes Énergies (LPTHE), UMR 7589, Sorbonne Université et CNRS, 4 place Jussieu, 75252 Paris Cedex 05, France\label{add1} \and\
  KEK Theory Center, Tsukuba 305-0801, Japan\label{add2} \and\
  Faculty of Science, Xi'an University of Architecture and Technology, Xi'an, 710055, China\label{add5} \and\
  Faculty of Education, Iwate University, Morioka, Iwate 020-8550, Japan\label{add6}
}

\date{\today}
\maketitle

\vspace*{-8.0cm}
  \noindent {\small\texttt{KEK-TH-2668}}
\vspace*{7.0cm}

\begin{abstract}
  We present a method to simulate toponium formation events at the LHC using the Green's function of non-relativistic QCD in the Coulomb gauge, which governs the momentum distribution of top quarks in the presence of the QCD potential. This Green's function can be employed to re-weight any matrix elements relevant for $t\bar{t}$ production and decay processes where a colour-singlet top-antitop pair is produced in the $S$-wave at threshold. As an example, we study the formation of $\eta_t$ toponium states in the gluon fusion channel at the LHC, combining the re-weighted matrix elements with parton showering.
\end{abstract}


\section{Introduction} \label{sec:intro}
The formation of toponium, a colour-singlet bound state of top and antitop quark in which the constituent top and antitop decay into a $b$-quark and a $W$ boson, was predicted by Fadin and Khoze~\cite{Fadin:1987wz} many years before the top quark was discovered in 1995~\cite{CDF:1995wbb, D0:1995jca}. The absence of a top quark signal at LEP indicated that the top quark mass should exceed 100 GeV, and that it decays primarily into a $b$-quark and a $W$ boson. This ensures that the top decay width $\Gamma_t$ had to be of the order of 1~GeV. As a result, toponium systems decay instantly and cannot produce sharp resonance peaks as were seen for charmonium and the bottomonium states. The pioneering work in~\cite{Fadin:1987wz} showed that by using the Green's function of the non-relativistic QCD Hamiltonian evaluated at $E+i\Gamma_t$, we can capture the effects of toponium on the total cross section for the $e^+ e^- \to t \bar{t} \to b W^+ \bar{b} W^-$ process, and this was similarly addressed for gluon fusion process relevant for hadron colliders in~\cite{Fadin:1990wx,Hagiwara:2008df}.

These previous studies of toponium formation at the LHC indicated that its impact on the total top-antitop production cross section would be less than 1\%. As a result, the prevailing assumption was that these effects would be experimentally insignificant, and no further detailed investigations were pursued. However, motivated by a $3\sigma$ excess of di-leptonic top-antitop events in the smallest relative angle bin reported by the ATLAS collaboration~\cite{ATLAS:2019hau}, we showed in~\cite{Fuks:2021xje} that this observed excess is consistent with the expectation from pseudo-scalar toponium formation at the LHC~\cite{Hagiwara:2008df,Sumino:2010bv}. More recently, high-statistics measurements by the LHC collaborations~\cite{ATLAS:2023gsl, CMS:2024ybg} have revealed additional features in top-antitop events that could be consistent with toponium formation.

Our previous study~\cite{Fuks:2021xje} was achieved through leading-order (LO) simulations matched with parton showers (PS), using a simplified model where toponium was treated as a pseudo-scalar resonance coupling to pairs of gluons and top-antitop quarks. The obtained corrections, of a non-perturba\-tive origin, could then be added to the conventional Standard Model (SM) predictions in order to increase the accuracy of top-antitop production modelling at the LHC. In this report, we propose a strategy that does not require introducing any fictitious new particle beyond the SM. Instead, we focus on the production and decay of a colour-singlet top-antitop pair and re-weight the corresponding matrix elements using Green's function ratios to capture the leading toponium effects. These re-weighted matrix elements can be used for fixed-order event generation, matched with parton showers once the radiation pattern expected by a toponium state is incorporated at the hard-scattering level, and finally combined with conventional SM top-antitop events to get predictions for top-antitop production including the crucial non-perturbative contributions neglected so far.

The remainder of this study is organised as follows. In section~\ref{sec:Green}, we outline how to modify matrix elements relevant to the production of a colour-singlet top-antitop quark pair in hadronic collisions to account for toponium contributions through Green's function ratios. Section~\ref{sec:mg5} details the implementation of these modifications in the \mgamc event generator~\cite{Alwall:2014hca}, enabling fixed-order predictions matched with parton showers to supplement conventional SM top-antitop predictions, which typically neglect non-perturbative contributions. In sections~\ref{sec:eta_prod} and \ref{sec:lops}, we apply this strategy to study a few distributions highlighting the impact of the toponium effects in LHC collisions at a centre-of-mass energy of $\sqrt{s} = 13$~TeV, first at the level of hard-scattering events and next at LO+PS. Finally, in section~\ref{sec:conclusion}, we summarise our findings and provide an outlook on further improvements to enhance the accuracy of SM top-antitop production predictions.

\section{Non-relativistic QCD Green's function for toponium formation} \label{sec:Green}

We consider a generic hadron collider process where a top-antitop pair is produced at a specific point $x$ in position space. The produced top and antitop quarks then decay at positions $y$ and $z$, respectively. In the absence of QCD interactions between the top and antitop quarks, the three-point function $K(x,y,z)$, given by~\cite{Sumino:1992ai}
\be\label{eq:3point}
  K_{abcd}(x,y,z) = \big\langle 0 \big |\,  T\big\{t_c(y) \bar{t}_d(z) \!:\!  \bar{t}_a(x) t_b(x) \!:\! \big\} \, \big| 0 \big\rangle\,,
\ee
where the indices $a$, $b$, $c$ and $d$ are four-component spinor indices, simplifies to the product of free propagators for the top and antitop quarks. However, near the threshold (\ie\ when the produced top and antitop quarks have small relative velocities) and in the presence of QCD interactions, the $t\bar t$ system is non-relativistic and bound by the static QCD potential $V_\mathrm{QCD}(\vec{r})$. This regime allows for the formation of toponium resonances whose impact must be assessed. At a given time $x^0$, the $t\bar t$ system behaves as a plane wave packet, which expands until it reaches the QCD potential barrier at a distance given by the Bohr radius $a_0 = (C_F \alpha_s m_t/2)^{-1}$, with $C_F=4/3$ and where the $m_t/2$ factor stems from the reduced mass of the top-antitop system. Since the top mass $m_t \simeq 173$~GeV, the toponium size satisfies $a_0 \ll \Gamma_t^{-1}$, allowing the plane waves to oscillate within the potential before the top and antitop decay. Therefore, the system probes the QCD potential, and toponium effects must be accounted for in top-antitop production processes near threshold. In this non-relativistic regime, the Schrödinger equation can be used to estimate the impact of the strong interaction on the three-point Green's function $K(x,y,z)$.

By separating the time and spatial dependence of the spinor fields and using the completeness of the basis of $|\vec{r}\rangle$ states, we can express \eqref{eq:3point} as
\be\label{eq:3point_exp}\begin{split}
  &K_{abcd}(x,y,z) = \frac{(1+\gamma^0)_{ca}}{2}\, \frac{(1-\gamma^0)_{bd}}{2}\\
  &\quad \times \int \mathrm{d}^3r \Big[ K_1\big(y; (z^0, \vec{r})\big)\,  K_2(z^0, \vec{r}, \vec{z}; x^0, \vec{x}, \vec{x}) \\
  &\qquad\qquad\qquad + K_1\big(z; (y^0, \vec{r})\big) \, K_2(y^0, \vec{y}, \vec{r}; x^0, \vec{x}, \vec{x}) \Big]\,,
\end{split}\ee
where $(1+\gamma^0)/2$ and $(1-\gamma^0)/2$ are the standard non-relativistic projection operators for the top and antitop quarks, respectively. The two terms in the integrand represent the only two relevant time-ordered configurations for top-antitop production in the non-relativistic approximation. The first term corresponds to a setup where the toponium state forms at time $x^0$ and position $\vec{x}$, propagates as a two-particle state until time $z^0$ and positions $\vec{r}$ (for the top) and $\vec{z}$ (for the antitop). There, the antitop quark decays, and the top quark then propagates further until time $y^0$ and position $\vec{y}$ where it decays, leading to the time ordering $x^0 < z^0 < y^0$. The second term describes a reversed sequence where the toponium system propagates until time $y^0$ and positions $\vec{r}$ (for the antitop) and $\vec{y}$ (for the top). At this time, the top quark decays while the antitop quark propagates to time $z^0$ and position $\vec{z}$ where it decays. In this case, the time ordering is $x^0 < y^0 < z^0$. Thus, the three-point Green's function in \eqref{eq:3point_exp} can be expressed as a sum of two terms, each containing a two-particle-state propagator $K_2$ and a one-particle-state propagator $K_1$. Here, $K_1(y; x)$ represents the propagator of a free top quark from the space-time point $x$ to the space-time point $y$ (with $y^0>x^0$),
\be
  K_1(y; x) = \int \frac{\mathrm{d}^4p}{(2\pi)^4} \, \frac{i}{p^0 - m_t - \frac{|\vec{p}|^2}{2 m_t} + i\varepsilon} \, e^{-i p \cdot (y-x)}\,,
\ee
while $K_2(y^0, \vec{y}_1, \vec{y}_2; x^0, \vec{x}_1, \vec{x}_2)$ represents the two-particle-state propagator from time $x^0$ and positions $\vec{x}_1$, $\vec{x}_2$ to time $y^0$ and positions $\vec{y}_1$, $\vec{y}_2$. By splitting this two-particle propagator into components along the centre-of-mass coordinate, from $x_G\equiv (x^0, \vec{x}_G)$ to $y_G\equiv (y^0, \vec{y}_G)$, and the relative coordinate from $x_r\equiv (x^0, \vec{x}_r)$ to $y_G\equiv (y^0, \vec{y}_r)$, we can write
\be
  K_2(y^0, \vec{y}_1, \vec{y}_2; x^0, \vec{x}_1, \vec{x}_2) = K_G(y_G; x_G)\, K_r(y_r; x_r)\,,
\ee
where the two propagators in the right-hand side vanish for $y^0<x^0$. In this notation, the propagator $K_G$ describes the free propagation of the toponium's centre-of-mass and is given by
\be
  K_G(y; x) = \int \frac{\mathrm{d}^4p}{(2\pi)^4} \, \frac{i}{p^0 - 2m_t - \frac{|\vec{p}|^2}{4 m_t} + i\varepsilon} \, e^{-i p \cdot (y-x)}\,.
\ee
Meanwhile, $K_r$ is the kernel of the time-dependent Schrödinger equation,
\be
  \Big[ -\frac{\vec{\nabla}^2}{m_t} + V_\mathrm{ QCD}(\vec{x}-\vec{y}) - i\frac{\partial}{\partial t}\Big] K_r(y;x) = -i \delta^{(4)}(x-y)\,,
\ee
derived from the Hamiltonian of non-relativistic QCD in the Coulomb gauge, and with all derivatives taken with respect to the $x$ coordinates. Applying the time-derivative operator and introducing the width $2\Gamma_t$ of the two-particle toponium state, we obtain an equation for the associated Green's function $G(E; \vec{x})$ in position space~\cite{Fadin:1987wz, Strassler:1990nw, Sumino:1992ai},
\be\label{eq:xspaceGreen}
  \Big[ -\frac{\vec{\nabla}^2}{m_t} + V_\mathrm{QCD}(\vec{x}) -(E + i \Gamma_t) \Big] G(E; \vec{x}) = \delta^{(3)}(\vec{x}) \,.
\ee
Here, $E$ is the sum of the kinetic and potential energies of the top quark with reduced mass of $m_t/2$. This gives the toponium binding energy in the toponium rest frame, $E = E_t + E_{\bar{t}} - 2m_t$, with $E_t$ and $E_{\bar{t}}$ being the top and antitop energies in that frame. It has been shown~\cite{Fadin:1987wz, Fadin:1990wx} that the imaginary part of the Green's function at the origin, ${\rm Im}\, G(0;E+i\Gamma_t)$, provides the toponium contribution to any relevant scattering amplitude, higher-order corrections having been computed in more recent works~\cite{Kiyo:2008bv,Ju:2020otc}.

Taking the Fourier transform of~\eqref{eq:3point_exp}, we obtain an expression for the three-point function in the momentum-space, $\widetilde{K}(p_t, p_{\bar{t}})$, which depends on the four-momenta $p_t$ and $p_{\bar{t}}$ of the top and antitop quarks in the rest frame of the toponium system. Introducing $p^*$ as the common magnitude of the momentum $\vec{p}_t$ and $\vec{p}_{\bar{t}}$ in the top-antitop rest frame, we can write
\be\label{eq:momentum_3pt}\begin{split}
  &\widetilde{K}_{abcd}(p_t, p_{\bar{t}}) = \frac{(1+\gamma^0)_{ca}}{2}\, \frac{(1-\gamma^0)_{bd}}{2} \\ &\hspace{3.25cm}\times \widetilde{G}(E; p^*) \big[ D(p_t) + D(p_{\bar{t}})\big]\,,
\end{split}\ee
where $D(p)$ corresponds to the non-relativistic propagator for a free top (or antitop) quark in the momentum-space, given by
\be
  D(p) = \frac{i}{p^0 - m_t - \frac{|\vec{p}|^2}{2 m_t} + \frac{i}{2} \Gamma_t}\,.
\ee
In this expression,  we have explicitly accounted for the the top quark width $\Gamma_t$. Additionally, \eqref{eq:momentum_3pt} depends on $\widetilde G(E; p)$ that represents the Fourier transform of the toponium Green's function $G(E; \vec{x})$ defined in \eqref{eq:xspaceGreen}. This Green's function in the momentum-space can be obtained by solving the Lippmann-Schwinger equation, which depends on the Fourier transform of the QCD potential, $\widetilde{V}_\mathrm{QCD}(\vec{p})$~\cite{Jezabek:1992np, Hagiwara:2016rdv, Hagiwara:2017ban, Fuks:2021xje},
\be\label{eq:LS}
  \widetilde G(E; p) = \widetilde G_0(E; p) + \int \frac{\mathrm{d}^3q}{(2\pi)^3}\, \widetilde V_\mathrm{QCD}(\vec{p}-\vec{q})\, \widetilde G(E; q)\,.
\ee
Here $\widetilde G_0(E; p)$ is the Green's function for the free Hamiltonian with an energy shift to incorporate the top quark width effects, that is given by
\be\label{eq:freeG0}
  \widetilde{G}_0(E; p) = \frac{1}{(E+i\Gamma_t) - \frac{|\vec{p}|^2}{m_t}}\,.
\ee

At short distances and for a toponium system produced in the S-wave, one-gluon exchanges between the top and antitop quarks dominate so that the potential is perturbative and behaves like the Coulomb potential near the origin. In the momentum space, this results in the two-loop potential which reads, in the $\overline{\text{MS}}$ scheme,
\be\label{eq:coulomb}
  \widetilde{V}_\mathrm{Coul}(q^2) = -C_F \frac{4\pi\alpha_s(q^2)}{q^2} \bigg[ 1 + \Big(\frac{31}{3} - \frac{10}{9} n_F\Big) \frac{\alpha_s(q^2)}{4\pi}\bigg]\,,
\ee
where $n_F=5$ is the number of massless quark flavours. This Coulomb-like potential is advantageous because the Lippmann-Schwinger equation~\eqref{eq:LS} can be solved analytically when $\alpha_s$ is taken to be a constant, which then yields a Whittaker function. This contrasts with the QCD potential~\eqref{eq:coulomb} with a running coupling constant, for which \eqref{eq:LS} has to be solved numerically with standard methods~\cite{Jezabek:1992np}.

As an example, in our previous work~\cite{Fuks:2021xje} we adopted a tree-level Coulombic potential including a strong coupling constant normalised from its value at the $Z$-pole. This choice for the potential provides a good starting point for studying toponium formation, as it allows for analytical control of the results by freezing the value of the coupling constant at a specific scale. Additionally, it serves as a useful cross check of our numerical code, which is general and can handle any form of the QCD potential $\widetilde V_\mathrm{QCD}(\vec{p})$ in the momentum space. We determined the Green's function $\widetilde G(E; p)$ for the tree-level part of the QCD potential~\eqref{eq:coulomb} with a running strong coupling constant normalised at $\alpha_s(m_Z) = 0.12$, and a top quark mass $m_t = 173$~GeV and width $\Gamma_t=1.49$~GeV.\footnote{In order to avoid numerical issues caused by the infrared pole of the QCD running coupling, we applied a Richardson-like regularisation~\cite{Richardson:1978bt}, following the methodology outlined in \cite{Jezabek:1992np}. Thanks to the physical infrared cutoff provided by the top quark width $\Gamma_t\approx 1.5$~GeV, the results show minimal sensitivity to the long distance behaviour of the QCD potential.} The results are given in two data files, \lstinline{swData.M20.M5.txt} and \lstinline{swData.M5.P20.txt}, which contain tabulated information on ratios of the Green's functions $\widetilde{G}(E; p)/\widetilde{G}_0(E; p)$. This information is provided in the form of two-dimensional grids in $(E, p)$. The first file focuses on low binding energy values, containing 150 points for $E$ in the range $-20$~GeV $\leq E \leq -5$~GeV, and 400 points for $p$ in the range $0<p\leq 100$~GeV. The second file addresses larger binding energy values, with 1250 points for $E$ in the range $-5$~GeV $\leq E \leq 20$~GeV, and the same 400 points for $p$. Each line in the files corresponds to specific $E$ and $p$ values, and contains four entries, the binding energy $E$, the momentum $p$ and the real and imaginary part of the ratio of the Green's functions. These files are available from a public repository at \url{https://github.com/BFuks/toponium.git}. 

\begin{figure*}
  \centering
  \includegraphics[width=0.99\textwidth]{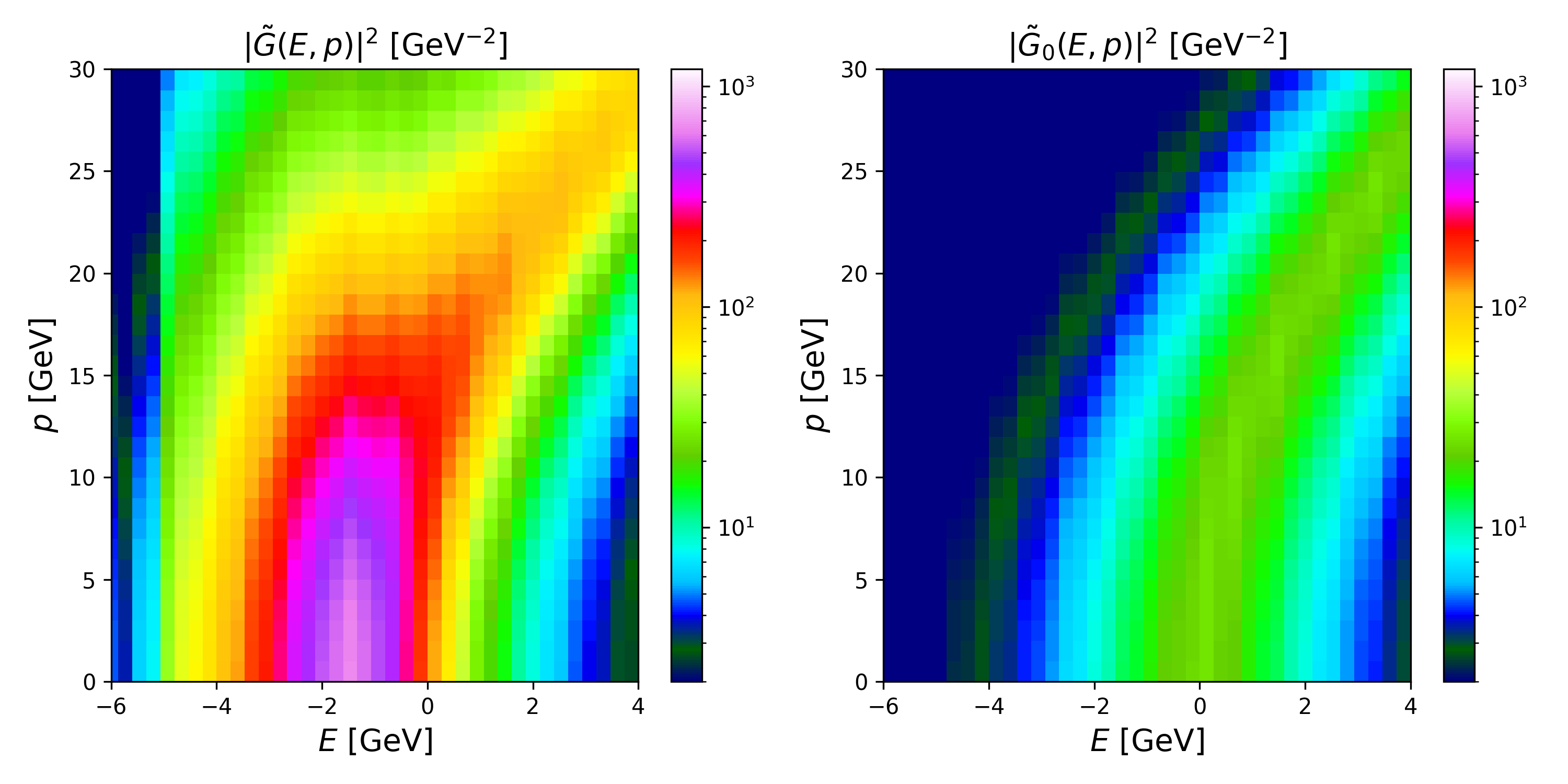}
  \caption{Norm of the Green's function $|\widetilde{G}(E; p)|^2$ in the presence of the tree-level Coulombic potential as obtained after fixing the strong coupling constant at the $Z$-pole $\alpha_s(m_Z)=0.12$ (left), and norm of the free Green's function~\eqref{eq:freeG0}, $|\widetilde{G}_0(E; p)|^2$ (right). The results are shown as a function of the binding energy $E$ and the momentum $p$.\label{fig:green}}
\end{figure*}
As an illustration, the left panel of figure~\ref{fig:green} shows the dependence of the Green's function squared norm $|\widetilde{G}(E; p)|^2$ in the presence of the Coulomb potential~\eqref{eq:coulomb}, as a function of the binding energy $E$ and the momentum $p$. In contrast, the right panel presents the norm of the free Green's function $|\widetilde{G}_0(E; p)|^2$. We observe notable differences between the two distributions, not only in magnitude but also in their overall shape, with deviations extending both above and below the threshold. In the free theory, $|G_0(E; p)|^2$ is given by \eqref{eq:freeG0}, and it should thus resembles a Breit-Wigner distribution for any specific fixed value of the binding energy $E$. This behaviour is the one depicted in the figure, with the position of the peak in $p$ depending on $E$ and being given by 
\be
  p_\mathrm{peak} \simeq \frac{2m_t+E}{2} \sqrt{1-\frac{4m_t^2}{(2 m_t +E)^2}}\,,
\ee
for $m_t=173$~GeV. This behaviour is approximately recovered for the interacting Green's function at positive binding energies ($E \gtrsim 2$~GeV), though with a different overall normalisation that persists until significantly larger values of $E$ (not shown on the figure). On the other hand, for $p\lesssim 20$~GeV and $-4$~GeV $\lesssim E \lesssim 1$~GeV, the toponium wave function not only enhances the magnitude of the Green's function squared norm, but also substantially distorts its shape, with the most prominent effect occurring around $E \simeq -2$~GeV at low momentum $p$.

The tabulated ratio of Green's functions in Fourier space $\widetilde{G}(E; p)/\widetilde{G}_0(E; p)$ is the quantity required when adjusting matrix elements relevant to top-antitop production near threshold to account for toponium effects. For example, consider the process 
\be\label{eq:6bodyttbar}
  gg \to t\bar{t} \to b \ell^+ \nu_\ell \bar{b} \ell^{\prime-} \bar{\nu}_\ell'\,,
\ee
in the colour-singlet channel, and where the intermediate top and antitop quarks can either be on-shell or off-shell. The process~\eqref{eq:6bodyttbar} includes top and antitop quark decays (that we have chosen leptonic), which ensures the opening of the pertinent phase space region below the $t\bar{t}$ production threshold and the correct embedding of all spin correlations among the final-state particles. To properly capture the toponium effects described earlier in this section, the corresponding squared matrix element $|M|^2$ should be re-weighted as
\be\label{eq:reweighting}
    |M|^2 \quad \to \quad |M|^2 \left|\frac{\widetilde{G}(E; p^*)}{\widetilde{G}_0(E; p^*)}\right|^2\,,
\ee
where $E = W - 2m_t = m(b\bar{b} \ell \ell' \nu_\ell \bar{\nu}_\ell') - 2m_t$ represents the toponium binding energy (with $W$ being the invariant mass of the reconstructed top-antitop system from the six-body final state), and $p^*$ is the common magnitude of the top and antitop quark momenta in the toponium rest frame.\footnote{The relative momentum appearing in the Hamiltonian of a top quark with a reduced mass $m_t/2$ and in the rest frame of the other top quark corresponds to the common momentum $p^*$ of the top and antitop quarks in the toponium rest frame.} In~\cite{Fuks:2021xje}, we applied the re-weighting procedure~\eqref{eq:reweighting} within the \mgamc software~\cite{Alwall:2014hca} to obtain fixed-order predictions for toponium effects at hadron colliders matched with parton showers. This earlier procedure relied on the introduction of an intermediate pseudo-scalar resonance to model toponium contributions to top-antitop production. In section~\ref{sec:mg5}, we discuss how the Green's function re-weighting procedure can be used without the need for any intermediate state.

\section{Toponium production with \mgamc}\label{sec:mg5}

We consider toponium contributions to top-antitop production at hadron colliders, \ie\ the process~\eqref{eq:6bodyttbar} with final-state electrons and muons, and we describe how to employ the \mgamc event generator~\cite{Alwall:2014hca} to model them at LO+PS. Although such toponium contributions should be included in precision computations for top-antitop production in the SM, they are generally neglected, reducing thus the accuracy of the predictions. This section aims to propose a strategy to address this issue.

This is achieved by entering the following commands in the \mgamc command line interface:
\begin{lstlisting}
  generate g g > t t$\sim$ > w+ b w- b$\sim$, \
      w+ > l+ vl, w- > l- vl$\sim$
  output eta_production
\end{lstlisting}
where we enforce that all contributing diagrams feature two internal top quark propagators, that could exchange Coulombic gluons representative of toponium bound state formation. The files within the directory named \lstinline{eta_production} generated by \mgamc require adjustments to ensure that the intermediate top-antitop system, or equivalently the six-body final-state system $b \ell^+ \nu_\ell \bar{b} \ell^{\prime-} \bar{\nu}_\ell'$, is in a colour-singlet state. Additionally, non-perturbative Coulombic corrections must be incorporated by multiplying the squared matrix element $|\mathcal{M}|^2$ by the square of the ratio of the Green's functions of the non-relativistic Hamiltonian and the free Hamiltonian as shown in \eqref{eq:reweighting}.

The above modification of the matrix element is sufficient for a standalone study of toponium effects. However, in order to jointly study it against SM top-antitop production which ignores non-perturbative QCD corrections as outlined in this report, care must be taken to avoid double-counting of some contributions that are included both in the toponium contributions and the perturbative SM ones. This could be accounted for through a slightly different re-weighting,
\begin{equation}\label{eq:reweighting_minus1}
  |\mathcal{M}|^2 \to |\mathcal{M}|^2\ \Bigg( \bigg|\frac{\widetilde{G}(E; p^*)}{\widetilde{G}_0(E; p^*)}\bigg|^2 - 1 \Bigg)\,.
\end{equation}
Here, the $-1$ term cancels the LO contribution of the perturbative SM events. Whereas some double counting persists, it can be considered as beyond LO.

These adjustments are implemented by modifying two of the \textsc{Fortran} files generated by \mgamc. Firstly, the matrix element implementation within the \lstinline{MATRIX1} function, included in the file \lstinline{matrix1_orig.f} in the sub-folder \lstinline{SubProcesses/P1_gg_wpbwmbx_wp_lvl_wm_lvl}, undergoes modifications. At the beginning of this function, we specify the utilisation of a new module called \lstinline{GREENDATA}. This module, described below, provides the value of the ratio of Green's function appearing in eq.~\eqref{eq:reweighting} for a given toponium invariant mass $W$ and momentum $p^*$, from tabulated data. This is achieved by adding one line at the beginning of the function \lstinline{MATRIX1},
\begin{lstlisting}
    REAL*8 FUNCTION MATRIX1(P,NHEL,IC,IHEL)

  C BEGIN addition
    USE GREENDATA
  C END addition

    IMPLICIT NONE
\end{lstlisting}
Secondly, several new variables must be declared for the calculation of the toponium kinematics. One possibility is to declare these variables just before the declaration of the so-called local variables in the \lstinline{MATRIX1} function. This involves adding the following lines:
\begin{lstlisting}
    INTEGER IHEL

  C BEGIN addition: toponium kinematics
    REAL*8 PMOM(0:3, 3)
    REAL*8 GXOM, GYOM, GZOM, GEOM
    REAL*8 GMOM, GTHE, GPHI, GBET, GGAM, GMAS
    REAL*8 EXXX, MXXX
    COMPLEX*16 GREEN
  C END addition
  
  C     
  C     LOCAL VARIABLES 
\end{lstlisting}

Next, colour factor adjustments are necessary to ensure that the matrix element generated by \mgamc corresponds specifically to the production of a colour-singlet final state. This involves modifying the implementation of the colour factor, which initially reads as follows (still inside the \lstinline{MATRIX1} function):
\begin{lstlisting}[mathescape=false]
  C     COLOR DATA
  C     
    DATA (CF(I,1),I=1,2)/
   $ 5.333333333333333D+00,
   $ -6.666666666666666D-01/
  C     1 T(1,2,5,8)
    DATA (CF(I,2),I=1,2)/
   $ -6.666666666666666D-01,
   $  5.333333333333333D+00/
  C     1 T(2,1,5,8)
  C     ----------
\end{lstlisting}
This matrix includes both a colour-singlet and a colour-octet component. It is derived from the colour structure inherent in all diagrams related to the $2\to2$ process $g^ag^b\to t^m\bar{t}_n
$, which involve a two-dimensional colour basis $\mathcal{C}$ given by
\renewcommand{\arraystretch}{1.5}
\be
  \mathcal{C}^{abm}{}_n = \bpm (T^a T^b)^m{}_n\\  (T^b T^a)^m{}_n\epm\,.
\ee
In this notation, $a$ and $b$ represent colour-octet indices associated with the initial-state gluons $g_a$ and $g_b$, while $m$ and $n$ represent colour-triplet indices associated with the final-state top quark and antiquark, $t^m$ and $\bar{t}_n$. Here, we use upper indices for indices related to the fundamental representation of $SU(3)_c$ and lower indices for indices related to the anti-fundamental representation. Moreover, the matrices $T$ represent the generators of $SU(3)_c$ in the fundamental representation. The full colour matrix $\mathcal{M}_c$ associated with the process $gg\to t\bar{t}$ is then determined by
\be\setlength\arraycolsep{6pt}
  \mathcal{M}_c = \mathcal{C}\, \mathcal{C}^\dagger = \frac{1}{4 N_c} \bpm (N_c^2-1)^2 & 1-N_c^2\\ 1-N_c^2 & (N_c^2-1)^2 \epm \,,
\ee
where $N_c=3$ is the number of colours. However, only colour-singlet contributions are relevant for toponium production. Thus, we need to replace the full colour matrix $\mathcal{M}_c$ with a different colour matrix $\mathcal{M}_c^{(1)}$ that solely includes the singlet contributions. This is achieved by multiplying each element of the colour basis $\mathcal{C}$ by a colour-singlet projector operator,
\be\label{eq:singletprojector}
  \mathcal{P}_m{}^n = \frac{1}{\sqrt{N_c}} \delta_m{}^n\,, 
\ee
giving
\be\setlength\arraycolsep{6pt}\label{eq:singletmatrix}
  \mathcal{M}_c^{(1)} = \frac{1}{4 N_c} \bpm N_c^2-1 & N_c^2-1\\ N_c^2-1 & N_c^2-1 \epm \,.
\ee
Correspondingly, the implementation of the colour matrix in the function \lstinline{MATRIX1} that was provided above has to be replaced by:
\begin{lstlisting}[mathescape=false]
  C     COLOR DATA
  C BEGIN modifications 
    DATA (CF(I,1),I=1,2)/
   $ 6.666666666666666D-01,
   $ 6.666666666666666D-01/
  C     1 T(1,2,5,8)
    DATA (CF(I,2),I=1,2)/
   $ 6.666666666666666D-01,
   $ 6.666666666666666D-01/
  C     1 T(2,1,5,8)
  C END modifications 
\end{lstlisting}

Next, we need to evaluate the invariant mass $W$ of the top-antitop system, or correspondingly the value of the binding energy $E$, together with the recoil momentum $p^*$ of the top (or the antitop) in the toponium rest frame. We recall that the two quantities $E$ and $p^*$ are the input parameters relevant for the matrix-element re-weighting introduced in Eq.~\eqref{eq:reweighting}. This is achieved by first reconstructing the four-momentum of the top quark $p$, the four-momentum of the antitop quark $\bar{p}$ and the four-momentum of the toponium system $P$ in the laboratory frame, using the six final-state four-momenta generated by \mgamc. Correspondingly, in the generated code, prior to the calculation of the \lstinline{MATRIX1} variable, the following lines are added:
\begin{lstlisting}
  C BEGIN addition
    DO M = 0, 3
      PMOM(M, 2) = 0
      PMOM(M, 3) = 0
      DO I = 3, 5
         PMOM(M, 2) = PMOM(M, 2) + P(M, I)
         PMOM(M, 3) = PMOM(M, 3) + P(M, I+3)
      ENDDO
      PMOM(M, 1) = PMOM(M, 2) + PMOM(M, 3)
     ENDDO
    ...
  C END addition 
    MATRIX1 = 0.D0 
\end{lstlisting}
Here, the variables \lstinline{PMOM(M,1)}, \lstinline{PMOM(M,2)} and \lstinline{PMOM(M,3)} respectively denote the four-momenta of the toponium system, the intermediate top quark and the intermediate top antiquark (with the Lorentz index \lstinline{M} ranging from 0 to 3). They are reconstructed from the final-state four-momenta \lstinline{P(M, I)} with \lstinline{I=3,4,5} for the $b$, $\ell^+$ and $\nu_\ell$ particles in the process \eqref{eq:6bodyttbar}, and \lstinline{I=6,7,8} for the $\bar b$, $\ell^{\prime '}$ and $\bar\nu'_\ell$ particles in that process. The dots in the code (\lstinline{...}), that we specify below, include appropriate Lorentz transformations to evaluate $p^*$ in the toponium rest frame. These transformations involve a rotation of angle $\varphi$ along the $z$-axis, a rotation of angle $\theta$ along the $y$-axis, and finally, a boost of velocity $\beta$ (and associated Lorentz factor $\gamma$) along the $z$-axis. These parameters are determined from the toponium four-momentum in the laboratory frame $P = (E^{t\bar{t}}, \vec{p}^{t\bar{t}}) = (E^{t\bar{t}}, p_x^{t\bar{t}}, p_y^{t\bar{t}}, p_z^{t\bar{t}})$ and the invariant mass $W$ of the top-antitop system,
\be\label{eq:torestframe}\begin{split}
  &\cos\theta = \frac{p_z^{t\bar{t}}}{|\vec{p}^{t\bar{t}}|}\,,\qquad\quad
  \tan\varphi = \frac{p_y^{t\bar{t}}}{p_x^{t\bar{t}}}\,,\\
  &\beta = \frac{|\vec{p}^{t\bar{t}}|}{E^{t\bar{t}}}\,,\qquad\qquad\ \
  \gamma = \frac{E^{t\bar{t}}}{W}\,,
\end{split}\ee
which are implemented as follows:
\begin{lstlisting}[mathescape=false]
  C     Lorentz transformation factors
    GMOM = SQRT(PMOM(1,1)**2 + PMOM(2,1)**2 +
   $ PMOM(3,1)**2)
    GMAS = SQRT(PMOM(0,1)**2 - GMOM**2)
    GTHE = DACOS(PMOM(3,1)/(GMOM+1.e-8))
    GPHI = ATAN2(PMOM(2,1), PMOM(1,1))
    GBET = GMOM/PMOM(0,1)
    GGAM = PMOM(0,1)/GMAS
\end{lstlisting}
The corresponding transformation of the four-momenta $p$, $\bar{p}$ and $P$ are then implemented as:
\begin{lstlisting}
  C     Rotating around Z-axis
    DO M=1,3
      GXOM = PMOM(1,M)
      GYOM = PMOM(2,M)
      PMOM(1,M)=COS(GPHI)*GXOM+SIN(GPHI)*GYOM
      PMOM(2,M)=COS(GPHI)*GYOM-SIN(GPHI)*GXOM
    ENDDO
  C     
  C     Rotating around Y-axis
    DO M=1,3
      GXOM = PMOM(1,M)
      GZOM = PMOM(3,M)
      PMOM(1,M)=COS(GTHE)*GXOM-SIN(GTHE)*GZOM
      PMOM(3,M)=COS(GTHE)*GZOM+SIN(GTHE)*GXOM
    ENDDO
  C     
  C     Boost along Z-axis
    DO M=1,3
      GEOM = PMOM(0,M)
      GZOM = PMOM(3,M)
      PMOM(0,M)=GGAM*GEOM-GGAM*GBET*GZOM
      PMOM(3,M)=GGAM*GZOM-GGAM*GBET*GEOM
    ENDDO
\end{lstlisting}

The code is now equipped to implement the re-weighting of Eq.~\eqref{eq:reweighting}.\footnote{Modifications to accommodate the re-weighting prescription of Eq.~\eqref{eq:reweighting_minus1} are trivial to implement, and are thus not further discussed.} This involves incorporating the ratio of Green's function $|\widetilde{G}(E, p^*)/\widetilde{G}_0(E, p^*)|^2$ into the computation immediately after the evaluation of the squared matrix element $|\mathcal{M}|^2$ (stored in the variable \lstinline{MATRIX1}). This is achieved by adding the following lines:
\begin{lstlisting}[mathescape=false]
  C BEGIN addition
  C     Inclusion of the Green`s function
    EXXX = PMOM(0, 1) - 2.0 * 173.D0
    MXXX = SQRT(PMOM(1,2)**2 + PMOM(2,2)**2
   $ + PMOM(3,2)**2)

    CALL CALCGREEN(EXXX,MXXX,GREEN,EDATA,
   $ PDATA,GRATA,GIATA,EEATA,PPATA,GGRTA,
   $ GGITA)

    IF(GMAS.LE.350.D0 .AND. MXXX.LE.50.D0) THEN
      MATRIX1 = GREEN*DCONJG(GREEN)*MATRIX1
    ELSE
      MATRIX1 = MATRIX1*0.D0
    ENDIF
  C END addition

    IF(SDE_STRAT.EQ.1)THEN
\end{lstlisting}
This piece of code first calculates the binding energy $E$ (\lstinline{EXXX}) and the magnitude of the top quark recoil momentum $p^*$ (\lstinline{MXXX}) in the toponium rest frame, using the momenta previously calculated. These values serve as arguments for the \lstinline{CALCGREEN} subroutine, which computes the ratio of Green's functions $\widetilde{G}(E, p^*)/\widetilde{G}_0(E, p^*)$ and stores it in the variable \lstinline{GREEN}. Finally, the code implements re-weighting by multiplying the matrix element according to Eq.~\eqref{eq:reweighting} if $W<350$~GeV and $p^*<50$~GeV. Although values for the ratio $\widetilde{G}(E, p^*)/\widetilde{G}_0(E, p^*)$ are provided in an extended range of $E$ and $p^*$, we restrict the use of the re-weighting to a regime such that $E \leq 4$~GeV and $p^* < 50$~GeV in order to ensure non-relativistic kinematics. The impact of this choice will be commented in section~\ref{sec:eta_prod}, in which we assess the effects of allowing $p^*<100$~GeV.

Our modifications finally include the implementation of the \lstinline{CALCGREEN} subroutine, which is added at the end of the file \lstinline{matrix1_orig.f}. This is given by:
\begin{lstlisting}[mathescape=false]
    SUBROUTINE CALCGREEN(EOMX,MOMX,GREEN,
   $ EDATA,PDATA,GRATA,GIATA,EEATASG,PPATASG,
   $ GGRTASG,GGITASG)

    COMPLEX*16 GREEN, IMAG1, ZREEN, NREEN
    PARAMETER (IMAG1=(0D0,1D0))

    INTEGER NMOM, NEOM, NMMM, NEEM, II, JJ
    PARAMETER (NMOM=400, NEOM=151, NMMM=400,
   $ NEEM=1251)
    REAL*8 DEOM,DEEM, DMOM,DMMM, EMIN,EMID,
   $ EMAX, MMAX, OREEN
    PARAMETER (DEOM=0.1,DEEM=0.02,DMOM=0.25,
   $ DMMM=0.25,EMIN=-20.,EMID=-5.,EMAX=20.,
   $ MMAX=100.)
    REAL*8,dimension(NEOM,NMOM)::EDATA,PDATA
    REAL*8,dimension(NEOM,NMOM)::GRATA,GIATA
    REAL*8,dimension(NEEM,NMMM)::EEATASG,
   $ PPATASG
    REAL*8,dimension(NEEM,NMMM) :: GGRTASG, 
   $ GGITASG

    REAL*8 EOMX, MOMX, AA, BB, RGR, IGR

    IF( EOMX .GT. EMIN .and. EOMX .LT. EMAX .and. MOMX .LT. 80 ) THEN
      ZREEN = (EOMX-MOMX**2./173.+IMAG1*1.49)
      IF(EOMX .LE. EMID) THEN
        II = (EOMX - EMIN)/DEOM
        IF(II .LT. 1) II = 1
        JJ = MOMX/DMOM
        IF(JJ .LT. 1) JJ = 1
        AA = (EOMX - EDATA(II,1))/DEOM
        BB = (MOMX - PDATA(1,JJ))/DMOM
        RGR = (1.-AA)*(1.-BB)*GRATA(II,JJ)+
   $          AA*(1.-BB)*GRATA(II+1,JJ)
        RGR = RGR+(1.-AA)*BB*GRATA(II,JJ+1)+
   $          AA*BB*GRATA(II+1,JJ+1)
        IGR = (1.-AA)*(1.-BB)*GIATA(II,JJ)+
   $          AA*(1.-BB)*GIATA(II+1,JJ)
        IGR = IGR+(1.-AA)*BB*GIATA(II,JJ+1)+
   $          AA*BB*GIATA(II+1,JJ+1)
        GREEN = (RGR + IMAG1 * IGR)*ZREEN
      ELSE
        II = (EOMX - EMID)/DEEM
        IF(II .LT. 1) II = 1
        JJ = MOMX/DMMM
        IF(JJ .LT. 1) JJ = 1
        AA = (EOMX - EEATASG(II,1))/DEEM
        BB = (MOMX - PPATASG(1,JJ))/DMMM
        RGR=(1.-AA)*(1.-BB)*GGRTASG(II,JJ)+
   $         AA*(1.-BB)*GGRTASG(II+1,JJ)
        RGR=RGR+(1.-AA)*BB*GGRTASG(II,JJ+1)+
   $         AA*BB*GGRTASG(II+1,JJ+1)
        IGR=(1.-AA)*(1.-BB)*GGITASG(II,JJ)+
   $         AA*(1.-BB)*GGITASG(II+1,JJ)
        IGR=IGR+(1.-AA)*BB*GGITASG(II,JJ+1)+
   $         AA*BB*GGITASG(II+1,JJ+1)
        GREEN = (RGR + IMAG1 * IGR)*ZREEN
      ENDIF
    ELSE
      GREEN = (1., 0.)
    ENDIF
    END
\end{lstlisting}
This subroutine interpolates tabulated values to obtain the ratio of Green's functions $\widetilde{G}(E, p^*)/\widetilde{G}_0(E, p^*)$ for specific binding energy $E$ in the range of $[-20, 20]$~GeV and recoil momentum $p^*<400$~GeV. These tabulated values, derived from the calculations described in section~\ref{sec:Green} are loaded at runtime from modifications made to the file \lstinline{driver.f}, located in the sub-folder \lstinline{SubProcesses/P1_gg_wpbwmbx_wp_lvl_wm_lvl}.

This relies on the module \lstinline{GREENDATA}, that was discussed earlier in this section, which includes all variables necessary to store the tabulated information on the ratio of Green's functions. The implementation of this module is performed at the beginning of the file \lstinline{driver.f}, as follows:
\begin{lstlisting}[mathescape=false]
    module GREENDATA
    implicit none

  C     Constants for the Green functions
    INTEGER NMOM,NEOM, NEEM,NMMM, IGRN,JGRN
    PARAMETER (NMOM=400, NEOM=151, NMMM=400, 
   $ NEEM=1251)    
    REAL*8,dimension(NEOM,NMOM)::EDATA,PDATA
    REAL*8,dimension(NEOM,NMOM)::GRATA,GIATA
    REAL*8,dimension(NEEM,NMMM)::EEATA,PPATA
    REAL*8,dimension(NEEM,NMMM)::GGRTA,GGITA
    COMMON EDATA, PDATA, GRATA, GIATA
    COMMON EEATA, PPATA, GGITA, GGRTA

    end module
\end{lstlisting}
The program \lstinline{DRIVER}, responsible for orchestrating the entire calculation in \mgamc, next undergoes modifications to integrate the functionality of the \lstinline{GREENDATA} module. This entails including the module at the beginning of the program, 
\begin{lstlisting}
    Program DRIVER
  C ***************************************
  C Driver for the whole calulation
  C ***************************************
  
  C BEGIN addition
    USE GREENDATA
  C END addition

    implicit none
\end{lstlisting}
as well as implementing the reading of the two files \lstinline{swData.M20.M5.txt} and \lstinline{swData.M5.P20.txt} discussed in section~\ref{sec:Green}. These files, available from the repository \url{https://github.com/BFuks/toponium.git}, must be copied into the sub-folder \lstinline{SubProcesses/P1_gg_wpbwmbx_wp_lvl_wm_lvl}. These files are read by adding, after the declaration of all variables in the program \lstinline{DRIVER}, the lines:
\begin{lstlisting}[mathescape=false]
  C  BEGIN CODE
  C----- 
  C 
  C BEGIN addition
    open(unit=99,file="../swData.M20.M5.txt")
    read(99,*) ((EDATA(IGRN,JGRN), 
   $  PDATA(IGRN,JGRN), GRATA(IGRN,JGRN),
   $  GIATA(IGRN,JGRN), JGRN=1, NMOM, 1),
   $  IGRN=1, NEOM, 1)
    open(unit=98,file="../swData.M5.P20.txt")
    read(98,*) ((EEATA(IGRN,JGRN), 
   $  PPATA(IGRN,JGRN), GGRTA(IGRN,JGRN),
   $  GGITA(IGRN,JGRN), JGRN=1, NMMM, 1),
   $  IGRN=1, NEEM, 1)
  C END addition

    call cpu_time(t_before)
 \end{lstlisting}

To prevent \mgamc from optimising the calculation and overriding the modifications made, it is essential to disable the automatic recycling of helicity amplitudes inherent to the code. This can be achieved by adding the following line at the end of the file \lstinline{run_card.dat}, located in the \lstinline{Cards} sub-folder of the working directory:
\begin{lstlisting}
  False = hel_recycling
\end{lstlisting}

Hard-scattering event generation in \mgamc follows the usual procedure, initiated from the working directory by typing in a shell the command
\begin{lstlisting}
  ./bin/generate_events
\end{lstlisting}
It is crucial to note that at this stage, only hard-scattering events can be reliably generated. Attempting to match them with parton showering using \mgamc's automated methods or directly generating LO+PS events would lead to incorrect results. This originates from the fact that by default, automatic LO+PS event generators assume that the intermediate virtual top and antitop quarks could radiate off gluons, even when they are the constituents of a colour-singlet toponium state. It is therefore necessary to ensure that the initial-state gluon pair and the final-state $b \ell^+ \nu_\ell \bar{b} \ell^{\prime-} \bar{\nu}_\ell'$ system, as well as the intermediate top-antitop system, are all colour singlets. This adjustment guarantees correct parton shower behaviour, including a large amount of initial-state radiation, only final-state radiation from the produced $b$ quark and antiquark, and no radiation from the intermediate top-antitop system. As discussed above, at the peak of the toponium contribution where $E \approx -2$~GeV or $W \approx 344$~GeV, the ground state toponium has a size of about $20$~GeV$^{-1}$ (the Bohr radius). Notably, this was verified by Green’s functions calculated using the QCD potential with NLO corrections. Consequently, no QCD radiation with energies below 20~GeV can be emitted from the toponium system.

This is achieved through a two-step procedure. Firstly, the colour-flow information of all coloured particles in the event is updated. This involves ensuring that the colour and anti-colour identifiers of the initial-state gluons complement each other. Similarly, it is essential that the unique colour identifier of the intermediate top quark and final-state bottom quark aligns with the anti-colour identifier of the intermediate top antiquark and final-state bottom antiquark. Additionally, to properly account for the colour-singlet nature of the intermediate $t\bar{t}$ system and the six-body final-state system $b\ell^+\nu_\ell\bar{b}\ell^-\bar{\nu}_\ell$ during the parton showering stage that we handle using \py~\cite{Bierlich:2022pfr}, an artificial intermediate $Z'$ resonance is introduced. This $Z'$ resonance is assigned a four-momentum matching that of the toponium state. By design, \py recognises this fake $Z'$ resonance as a colour-singlet state, thus preserving its invariant mass during showering. This preservation of invariant mass is crucial for accurate toponium predictions, as it plays a pivotal role in evaluating the ratio of Green's functions used to re-weight the process matrix element. Finally, it is possible that the intermediate top and antitop quarks are not written by \mgamc to the event record because they are too off-shell.\footnote{\mgamc only writes to the event file top quarks and antiquarks with an invariant mass lying in the range $[m_t-b_w \Gamma_t, m_t+b_w \Gamma_t]$. By default, the $b_w$ parameter, named \lstinline{bwcuttoff} in the \lstinline{run_card.dat} configuration file, is set to 15.} To address this, we utilise the information determined during the fake $Z'$ reconstruction procedure to re-add these missing particles in the event record.

After implementing the necessary modifications, a typical toponium event in the event file would appear as follows:
\begin{lstlisting}
  <event>
  13 1 +2.71e-01 1.74e+02 7.55e-03 1.08e-01
    21 -1 0 0 501 502 ...
    21 -1 0 0 502 501 ...
    32  2 1 2   0   0 ...
     6  2 3 3 503   0 ...
    24  2 4 4   0   0 ...
    -6  2 3 3   0 503 ...
   -24  2 6 6   0   0 ...
   -11  1 5 5   0   0 ...  
    12  1 5 5   0   0 ...  
     5  1 4 4 503   0 ...
    13  1 7 7   0   0 ...
   -14  1 7 7   0   0 ...
    -5  1 6 6   0 503 ...
  </event>
\end{lstlisting}
In this illustrative representation, we have omitted the momentum information for brevity and replaced it with ellipses. Additionally, we have limited, in the above illustration, the floating-point numbers to three digits in the first line of the event, which contains general information. Following the standard LHE format, each of the following lines within the event corresponds to a particle. The first entry indicates the particle identifier according to the Particle Data Group numbering scheme (\lstinline{21} corresponding to a gluon, \lstinline{6} to a top quark, \etc.), and the second entry refers to the initial-state (\lstinline{-1}), intermediate-state (\lstinline{2}) or final-state (\lstinline{1}) nature of the particle. The third and fourth entries provide information on the parent of each particle, while the fifth and sixth entries provide the colour and anti-colour information of each particle.

A \lstinline{Python} script called \lstinline{reprocess.py} incorporating these modifications is available on the repository \url{https://github.com/BFuks/toponium.git}, and should be run from any of the \lstinline{Events/run_xx} sub-folders generated by \mgamc. Once these modifications are done, parton showering can be simulated from the working directory by typing in a shell the command
\begin{lstlisting}
  ./bin/madevent 
\end{lstlisting}
This initiates the command line interface of \mgamc, from which we then type the command
\begin{lstlisting}
  pythia8 run_xx
\end{lstlisting}
where \lstinline{run_xx} indicates the relevant run name.

\section{Hard-scattering top-antitop production near threshold}\label{sec:eta_prod}
\begin{figure*}
  \centering
  \includegraphics[width=0.98\linewidth]{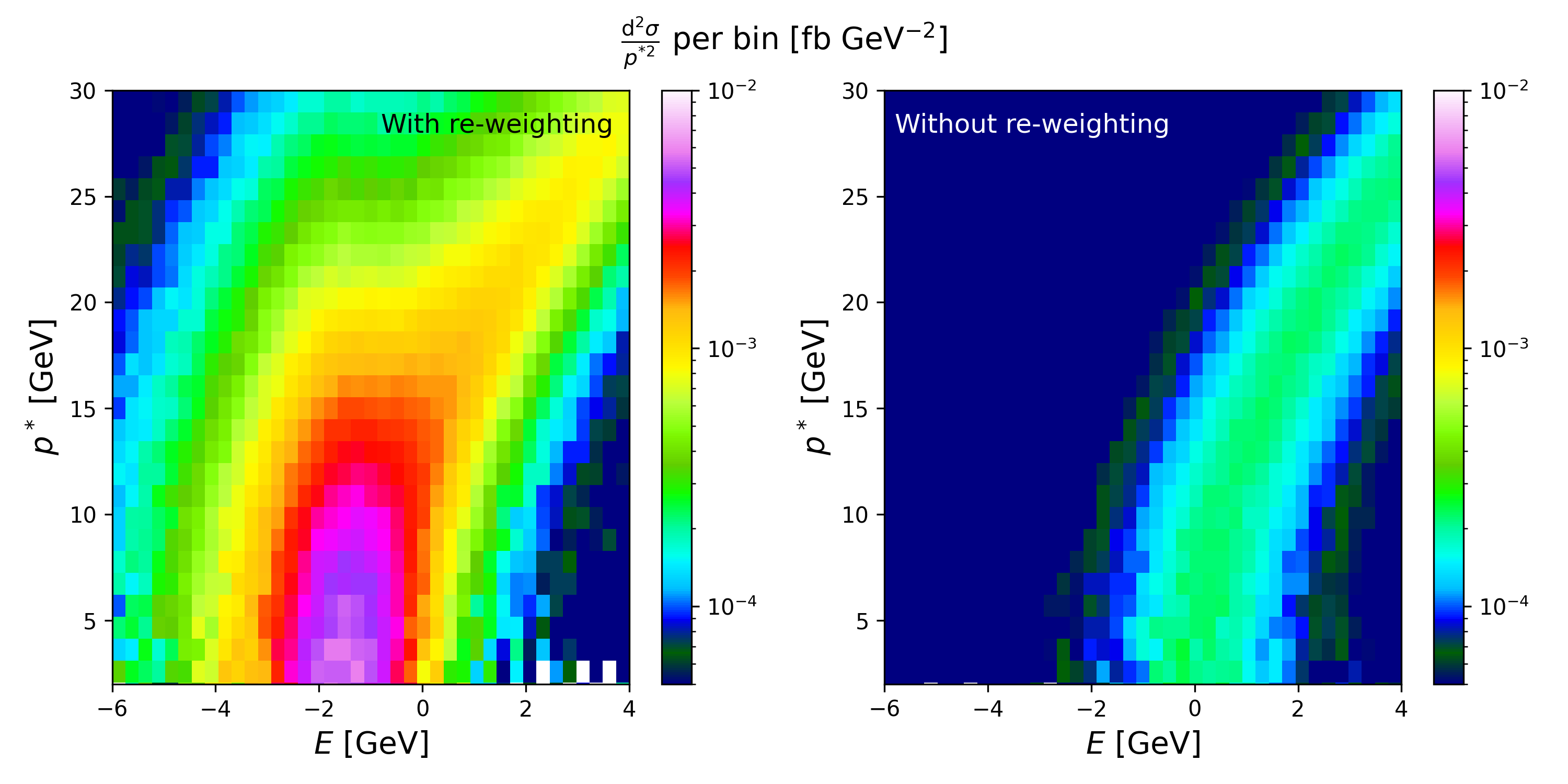}
  \caption{Two-dimensional differential cross section $\nicefrac{\mathrm{d}^2\sigma}{(p^*)^2 \mathrm{d}p^* \mathrm{d}E}$ for colour-singlet top-antitop production and di-leptonic decay as given by the process~\eqref{eq:6bodyttbar} in the top quark momentum magnitude $p^*$ in the toponium rest frame and binding energy $E$, at $\sqrt{s}=13$~TeV. We consider the two cases where the Green's function re-weighting is included (left panel) and without it (right panel).\label{fig:2D_Ep*}}
\end{figure*}

To assess the impact of toponium formation on the kinematic properties of the reconstructed top and antitop quarks, we follow the approach outlined in section~\ref{sec:mg5}. We begin by generating 1,000,000 hard-scattering events with \mgamc, simulating the production and decay of a colour-singlet top-antitop pair decaying di-leptonically (\ie\ corresponding to the process~\eqref{eq:6bodyttbar}) in proton-proton collisions at a centre-of-mass energy of $\sqrt{s} = 13$~TeV. We use the CT18NLO set of parton densities~\cite{Hou:2019efy}, that we access via \textsc{Lhapdf6}~\cite{Buckley:2014ana}, and we fix the top quark mass to $m_t=173$~GeV and top width to $\Gamma_t=1.49$~GeV. Moreover, we apply generator-level selection cuts to focus on the phase-space region defined by $340$~GeV $\leq W \leq 350$~GeV ($-6$~GeV $<E<4$~GeV) and $p^* < 50$~GeV. Simulations are performed both with and without Green's function re-weighting, which allows us to compare their respective effects after analysing the predictions by means of the \madanalysis software~\cite{Conte:2018vmg}. 

In figure~\ref{fig:2D_Ep*}, we present the doubly-differential cross section as a function of the binding energy $E$ and the common magnitude $p^*$ of the top and antitop momentum in the toponium rest frame. In the left panel of the figure, the (colour-singlet) matrix element includes the re-weighting~\eqref{eq:reweighting} by the ratio of Green's functions, while in the right panel of the figure it does not. The predictions shown in the right panel of figure~\ref{fig:2D_Ep*} are thus predominantly determined by the six-body phase space and a matrix element comprising Breit-Wigner propagators for the intermediate top and antitop quarks. This is confirmed by the shape of the heat map resulting from the \mgamc simulations, which exhibits the same two-dimensional structure expected from the dependence of the free Green's function on $E$ and $p^*$ shown in figure~\ref{fig:green} (right). As already manifest from the two-dimensional distributions exhibited by the free and interacting Green's functions, the re-weighting procedure~\eqref{eq:reweighting} therefore not only affects the rate, as evident from the different scales in the heat maps (\ie\ the colour code), but also significantly alters the shape of the distributions. The distribution in energy and momentum spanned by the free Green's function $|\widetilde G_0(E; p)|^2$ shown in figure~\ref{fig:green} (right) is indeed reproduced by the non-re-weighted event two-dimensional spectrum given in figure~\ref{fig:2D_Ep*} (right). Therefore, the spectrum emerging from re-weighted events in figure~\ref{fig:2D_Ep*} (left) directly measures the QCD Green's function $|\widetilde G(E; p)|^2$, which should be regarded as the momentum distribution of the top quark constituents in the toponium bound state. In other words, the structure of the toponium, the heaviest and smallest hadron of the SM, may be directly observable at the LHC. 

From the results of our simulations, we can determine that when the Green's function re-weighting is applied (\ie\ the distribution given in the left panel of figure~\ref{fig:2D_Ep*}), the typical top quark momentum,
\be
\big\langle p(E) \big\rangle  = \frac{\int \mathrm{d}^3p\, p \,\frac{\mathrm{d}^2\sigma}{p^2\, \mathrm{d}p\, \mathrm{d}E}}{\int \mathrm{d}^3p\, \frac{\mathrm{d}^2\sigma}{p^2\, \mathrm{d}p\, \mathrm{d}E}}\,,
\ee
is approximately 20~GeV for a binding energy of $E=-2$~GeV, which agrees with the prediction of non-relativistic QCD~\cite{Sumino:2010bv}. This value corresponds to the inverse of the Bohr radius of the toponium system, 
\be
  \frac{1}{a_0} = C_F\, \alpha_s(a_0^{-1})\, \frac{m_t}{2} \,,
\ee
where we assume a Coulombic approximation of the QCD potential with a constant value for the strong coupling $\alpha_s$. Using the estimate $a_0^{-1}\sim 20$~GeV, we find that the strong coupling at this scale is approximately $\alpha_s(a_0^{-1}) \sim 0.17$. This leads to a Coulomb potential strength of $-C_F \alpha_s(a_0^{-1})/{a_0} \sim -4.5$~GeV. Since the top quark’s kinetic energy is roughly $(p^*)^2/m_t \sim 2.3$~GeV, the total energy becomes approximately $-2.2$~GeV, which aligns well with the Green's function calculation using the QCD potential with a running coupling constant at two-loop accuracy.

\begin{figure}
  \centering
  \includegraphics[width=0.98\columnwidth]{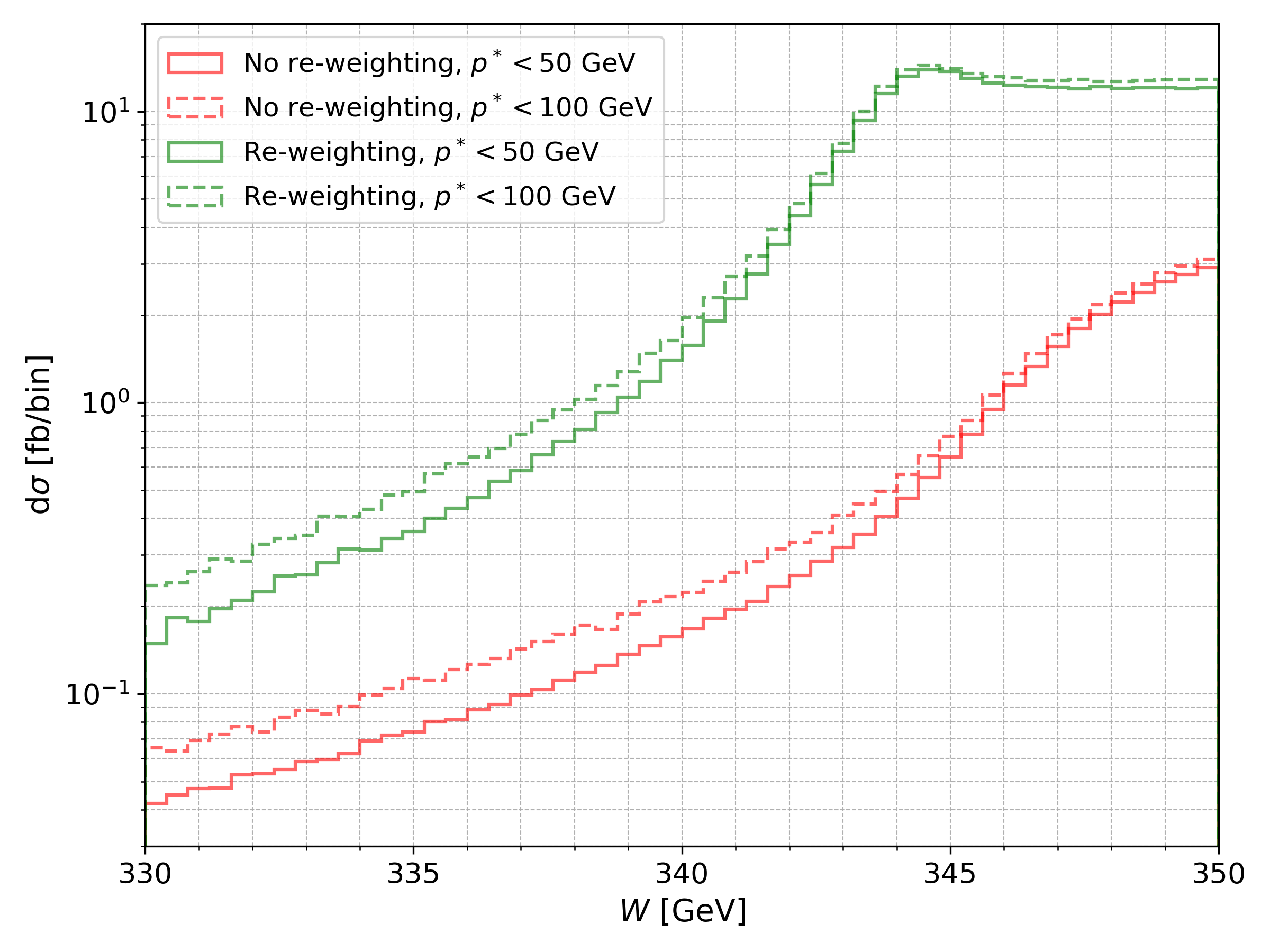}
  \caption{Distribution in the invariant mass $W\equiv m_{t\bar{t}}$ of a colour-singlet top-antitop system produced in proton-proton collisions at $\sqrt{s}=13$~TeV, when it decays di-leptonically (as described by the process~\eqref{eq:6bodyttbar}). We consider two cases, firstly when the Green's function re-weighting is included (solid, green) and secondly without it (solid, red). We additionally assess the impact of the requirement on $p^*$ by relaxing it to $p^*<100$~GeV (dashed curves).\label{fig:dsdW}}
\end{figure}
In figure~\ref{fig:dsdW}, we show the invariant mass distribution $\mathrm{d}\sigma/\mathrm{d}W$ of the reconstructed top-antitop system ($W = m_{t\bar{t}}$) that we obtain after integrating the previously discussed two-dimensional distribution over the top quark recoil momentum in the toponium rest frame (with $p^*<50$~GeV being enforced at the generator level). We therefore focus on the production of a colour-singlet top-antitop pair decaying di-leptonically, as defined in~\eqref{eq:6bodyttbar}. We compare results where the Green's function re-weighting~\eqref{eq:reweighting} is applied (green, solid) to those derived without it (red, solid). Both sets of predictions are in remarkable agreement with the potential non-relativistic QCD estimates of~\cite{Sumino:2010bv}, as well as with our previous modelling of toponium effects of~\cite{Fuks:2021xje}. We remind that the comparison with~\cite{Sumino:2010bv} should account for the top quark's leptonic branching ratio (around 20\%), the usage of a distinct set of parton densities, a larger value for the strong coupling constant at the $Z$ pole, and that we ignored colour-octet contributions that gain importance with increasing $W$ values. It is important to note that the method presented in section~\ref{sec:mg5} is not strictly non-relativistic, as the Green's function multiplies the full QCD matrix element instead of its non-relativistic counterpart, unlike in~\cite{Sumino:2010bv}. The validity of this approximation depends on the assumption that the full QCD matrix element deviates only minimally from its non-relativistic limit near the threshold. 

\begin{figure}
 \centering
  \includegraphics[width=0.98\columnwidth]{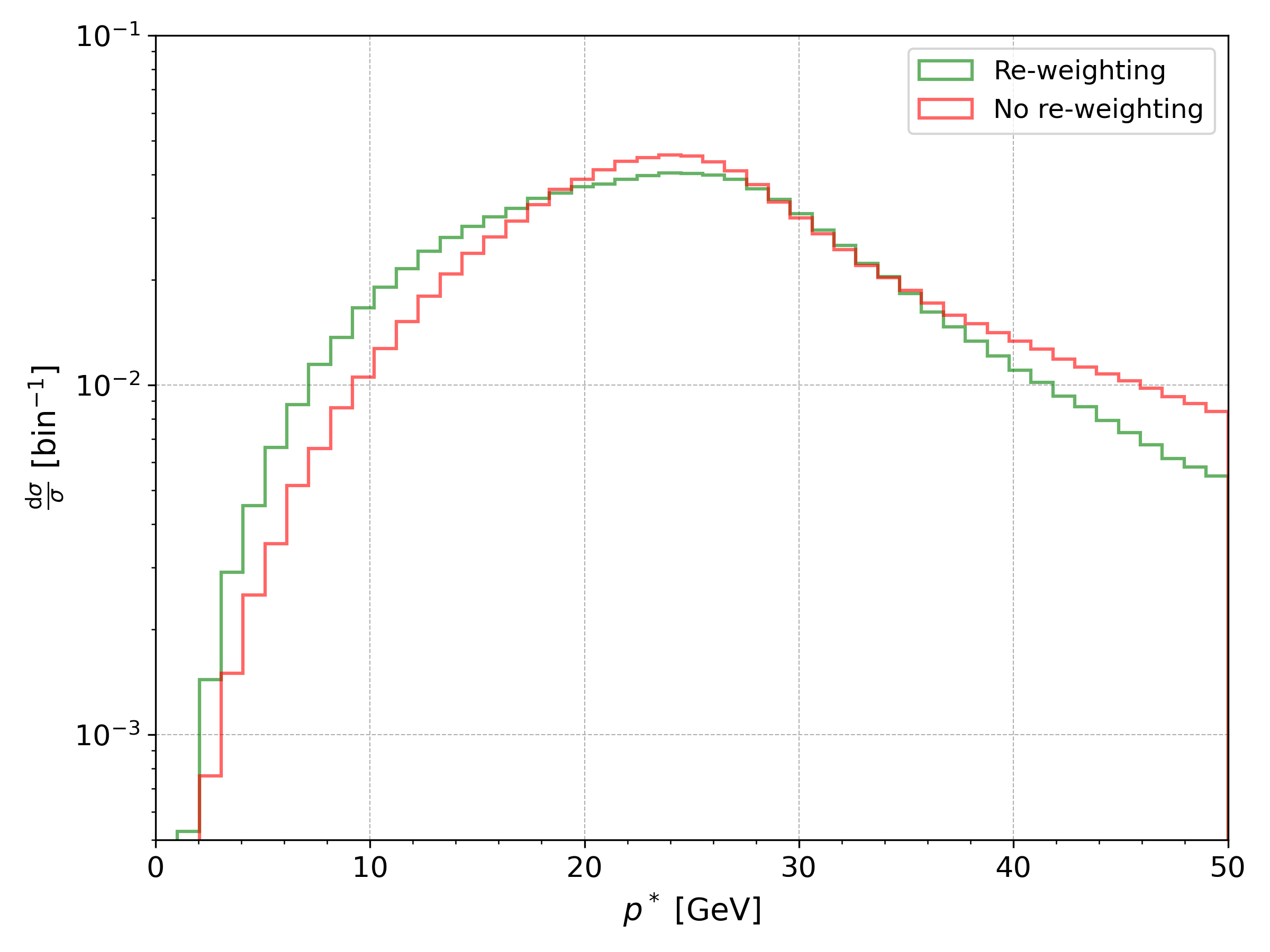} 
  \caption{Top recoil momentum distribution for colour-singlet top-antitop production without (red) and with (green) Green's function re-weighting. \label{fig:ds_dpcm}}
\end{figure}

\begin{figure*}
 \centering
  \includegraphics[width=0.98\textwidth]{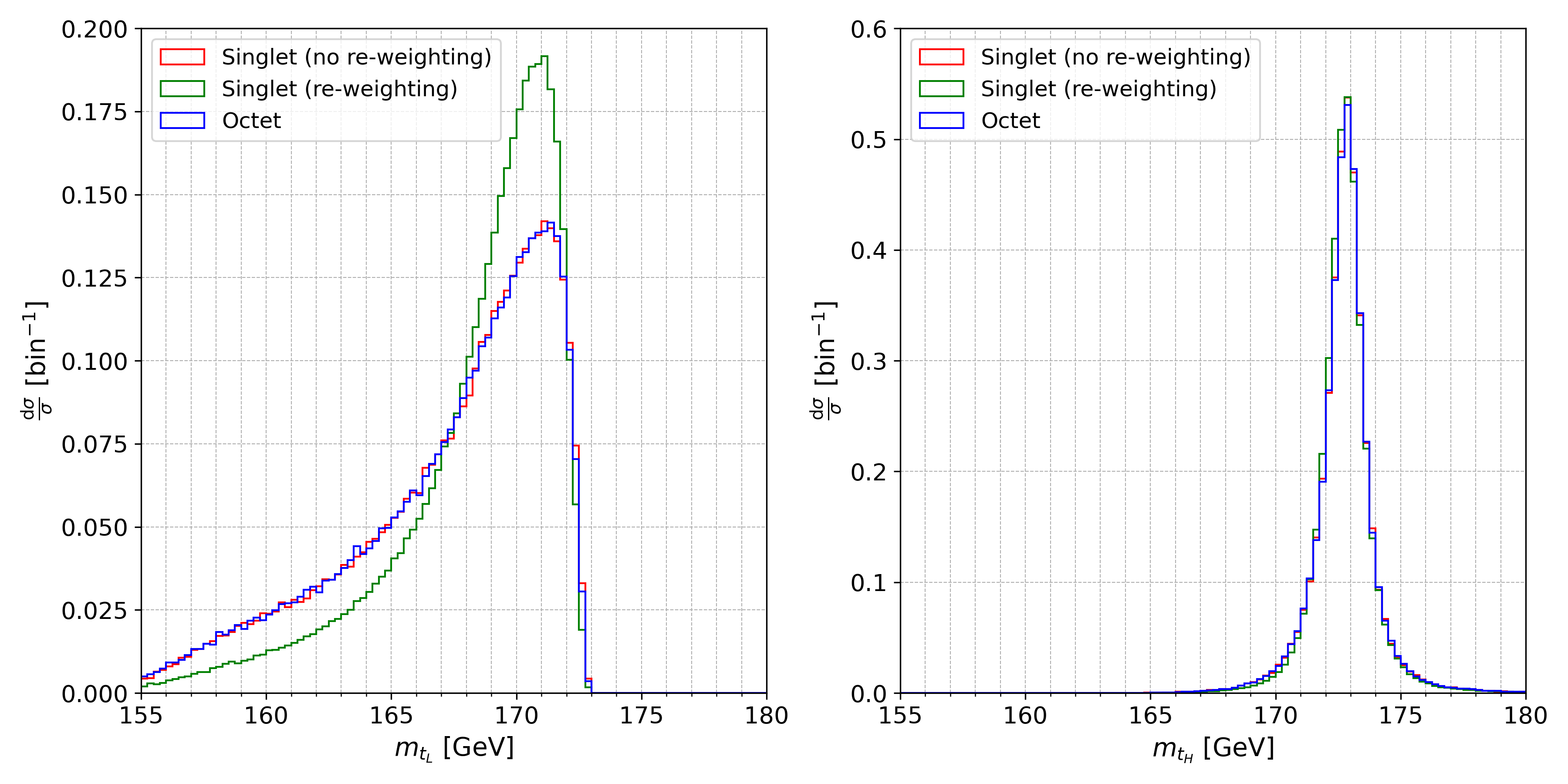} 
  \caption{Invariant mass distribution of the lighter reconstructed top quark ($m_{t_L}$, left) and heavier reconstructed top quark ($m_{t_H}$, right). The phase-space cuts $-4$~GeV $\leq E \leq 0$~GeV and $p^*\leq 50$~GeV are applied, and results are shown for colour-octet top-antitop production (blue), as well as colour-singlet top-antitop production without (red) and with (green) Green's function re-weighting.\label{fig:mt}}
\end{figure*}

As expected, toponium effects are strongest near $W\simeq 2m_t - 2$~GeV, with the observed broad peak corresponding to the impact of the $\eta_t$ pseudo-scalar toponium state. The influence of toponium contributions is found to extend beyond the peak, confirming the extended bound-state impact on the final-state kinematics. For higher $E$ values (not shown), the ratio of predictions converge to a constant, representing a $K$-factor that includes the higher-order contributions embedded in the calculation of $\widetilde{G}(E; p^*)$. However, we remind that toponium contributions as described by our modelling should not be included for $W>350$~GeV, as the average top quark velocity $\beta = 2 p^*/m_t$ becomes too large for a non-relativistic treatment to be valid. Figure~\ref{fig:dsdW} also examines the impact of relaxing the selection cut on $p^*$ to $p^*<100$~GeV through the two dashed curves. The effects are important far below the threshold, as could also be expected from figure~\ref{fig:2D_Ep*} that depicts a reduced but not negligible enhancement of the ratio of Green's functions with increasing $p^*$ values. For higher $W$ values, we start to probe a regime in which both the free Green's function and the interacting ones correspond to a Breit-Wigner, reducing thus the net impact of the cut on the recoil momentum $p^*$.

Since the $p^*$ distribution controls the structure of the toponium properties as a bound state, we show in figure~\ref{fig:ds_dpcm} the corresponding normalised distribution with (green) and without (red) the Green's function re-weighting. Without re-weighting, the distribution is determined by the phase space and the Breit-Wigner propagator of the top-quark. In the case of a modelling such that the total cross section would be artificially enhanced by the introduction a local pseudo-scalar boson that couples both to a pair of gluons and to a top-antitop pair, the obtained $p^*$ distribution should follow the non re-weighted predictions shown in the figure, since the associated tree-level matrix element for $gg\to t\bar t$ is essentially local near the threshold with a typical distance scale of $1/m_t$.

The $p^*$ distribution originating from toponium formation therefore subtly influences the kinematics of the six-body final state from the process~\eqref{eq:6bodyttbar}. This effect can be further illustrated by examining the invariant mass distributions of the two reconstructed top quarks $t_L$ and $t_H$ from the six-body final state. In this context, the invariant mass spectrum of the lighter top quark $t_L$ is of particular interest, as it is the first top quark to decay in the presence of the QCD potential of the remaining top quark. We present in figure~\ref{fig:mt} these invariant mass distributions, that we normalise to the total rate. We show predictions for both the lighter (left panel) and heavier (right panel) reconstructed top quarks, and for three different setups. We consider the production of a colour-singlet top-antitop pair with an unmodified matrix element (red curve), \ie\ without including the re-weighting factor~\eqref{eq:reweighting}, as well as the production of a colour-octet top-antitop pair (blue curve). For comparison, we superimpose to these predictions the invariant mass distributions obtained from the colour-singlet channel, but after re-weighting the corresponding matrix element by the ratio~\eqref{eq:reweighting} of Green's function (green curve). Furthermore, in each case we apply the phase space cuts $-4$~GeV $\leq E \leq 0$~GeV and $p^*\leq 50$~GeV. 

Starting with the distribution of the heavier top quark ($m_{t_H}$), we observe in all three considered cases a conventional Breit-Wigner shape with a peak at $m_t$ and a width given by $\Gamma_t$. This behaviour is not surprising, as the heavier quark $t_H$ decays after the lighter top quark $t_L$ has decayed. Thus, its invariant mass (in the case of the toponium signal) is not significantly affected by the bound-state dynamics, and remains close to the on-shell top mass. In contrast, the invariant mass distribution of the lighter top quark ($m_{t_L}$) is shifted to lower values when Green's function re-weighting is included. The Green's function $G(E; p^*)$ therefore governs the decay of the lighter top quark $t_L$, that is the first to decay within the Coulomb potential generated by the heavier top quark. In other words, in our formalism the lighter top quark is treated as a constituent bounded by the Coulomb potential generated by the heavy top quark, whereas the heavier quark can be seen as effectively stable until $t_L$ decays. As a consequence, the quark $t_L$ exhibits a momentum and invariant-mass that could be significantly altered relatively to the on-shell case, and when $t_H$ decays, its invariant mass distribution has a Breit-Wigner form, as this quark no longer experiences any bound-state formation effects. Strictly speaking, this picture is not entirely accurate, since the heavier top quark $t_H$ can still strongly interact with the $b$-quark originating from the decay of $t_L$ from deep inside the QCD potential (see \eg, \cite{Peter:1997rk}).

Finally, we observe that the results for the colour-octet and the unweighted colour-singlet production (thus without Green's function re-weighting) agree well, confirming that the dynamical effects introduced by the Green's function are indeed responsible for the observed differences in the $m_{t_L}$ distribution. These findings underscore the utility of the top quark reconstruction method proposed in~\cite{Fuks:2021xje}, and it suggests that the $p^*$ distribution can, in principle, be inferred from the shape of the $m_{t_L}$ distribution. Note that both the $m_{t_L}$ and $m_{t_H}$ distributions could in principle be measured directly in events where one of the top quarks decays hadronically. 

These results reinforce the space-time picture of toponium formation at the LHC sketched in section~\ref{sec:Green}. When a colour-singlet pair of gluons with the same helicity collides to produce a top-antitop quark pair, a toponium system with a size of $1/20$~GeV$^{-1}$ emerges at a time scale of $1/20$~GeV$^{-1}$, and subsequently decays at a time scale of $1/(2\Gamma_t) \sim 0.3$~GeV$^{-1}$. Both the formation and decay of the toponium occur much before hadronisation, which takes place at a time scale of $1$~fm (or $5$~GeV$^{-1}$). As a result, there exists a possibility to experimentally probe the toponium wave function, which may be regarded as the smallest hadron in the SM. It is however important to highlight a potential issue regarding the reconstruction of $b$ jets in toponium events, which could influence the invariant mass of the lighter top quark. At the peak of the toponium resonance, \eg\ at $E \approx -2$~GeV, the $b$ quark from the top quark that decays first is produced within a QCD potential of approximately $-4.5$~GeV. Heuristic studies~\cite{Peter:1997rk} suggest that the attractive interaction between the $b$-quark and the remaining top quark tends to reduce the value of $p^*$ and increase the value of $m_{t_L}$. The behaviour of $b$-jets emerging such deeply from the QCD potential has still to be thoroughly investigated, which goes beyond the scope of this work.

\section{Toponium phenomenology at the LHC}\label{sec:lops}
In this section, we employ the simulation chain introduced in section~\ref{sec:mg5} to generate 1,000,000 toponium events at LO+PS for proton-proton collisions at a centre-of-mass energy of $\sqrt{s} = 13$~TeV, either incorporating the re-weighting~\eqref{eq:reweighting} or not, together with the projection of the matrix element into a colour-singlet state. We use \mgamc~\cite{Alwall:2014hca} for hard-scattering simulations with the CT18NLO set of parton densities~\cite{Hou:2019efy, Buckley:2014ana}, and \py~\cite{Bierlich:2022pfr} for parton showering and hadronisation. The reconstruction of events is performed using the anti-$k_T$ algorithm~\cite{Cacciari:2008gp} with a jet radius parameter of $R = 0.4$, as implemented in \fastjet~\cite{Cacciari:2011ma} and interfaced with \madanalysis~\cite{Conte:2018vmg, Araz:2020lnp}. The latter software is also employed to apply the selection strategy previously introduced in \cite{Fuks:2021xje}. We then compare our predictions for the toponium signal with the background from conventional top-antitop pair production, followed by di-leptonic decays. We have simulated 3,100,000 background events at LO+PS accuracy, with a total cross section normalised to next-to-next-to-leading order with next-to-next-to-leading logarithmic threshold resummation (NNLO+NNLL)~\cite{Czakon:2013goa}, using thus a total cross section value of $\sigma_{t\bar{t}} = 36.9$~pb for the 2-to-6 scattering process.

We pre-select events that contain exactly two reconstructed $b$-jets and two reconstructed leptons (electrons or muons), each with a transverse momentum of $p_T > 25$~GeV and a pseudo-rapidity $|\eta| < 2.5$. Additionally, we require a transverse distance of $\Delta R > 0.4$ between the selected $b$-jets and leptons to ensure their spatial separation in the transverse plane. To improve the potential observation of the signal emerging from toponium decays, we impose constraints on the di-lepton system based on predictions from the toponium and top decay density matrices~\cite{Hagiwara:2017ban}. 

As done in \cite{Fuks:2021xje}, the properties of the angular distributions of di-leptons originating from toponium decays can be determined from the $t\bar t$ production density matrix,
\be
  \rho^{\sigma\bar\sigma,\sigma'\bar\sigma'}_{gg\to (t\bar t)_1} =  \frac{\sum\limits_{\lambda,\lambda'}{\cal M}^{\sigma\bar{\sigma}}_{\lambda\lambda'} \, \big({\cal M}^{\sigma'{\bar\sigma'}}_{\lambda\lambda'}\big)^*}{\sum\limits_{\sigma\bar{\sigma}} \sum\limits_{\lambda,\lambda'} \big|{\cal M}^{\sigma\bar{\sigma}}_{\lambda\lambda'}\big|^2}\,.
\ee
This density matrix is constructed from the matrix elements of the $gg\to t\bar t$ process in the colour-singlet channel, evaluated in the $gg$ (or $t\bar t$) rest frame, 
\be
  {\cal M}^{\sigma\bar{\sigma}}_{\lambda\lambda'} = {\cal M}\big(g_\lambda g_{\lambda'}\to t_\sigma \bar t_{\bar\sigma}\big)\,,
\ee
where $\sigma/2$ and $\bar\sigma/2$ denote the helicities of the top and antitop quarks, respectively, and $\lambda$ and $\lambda'$ denote the helicities of the initial-state gluons. The matrix elements in the colour-singlet channel are derived from the conventional matrix elements by inserting the projectors~\eqref{eq:singletprojector}, which leads to the colour structure~\eqref{eq:singletmatrix}. In this case, the only non-vanishing helicity amplitudes in the non-relativistic limit are
\be
  {\cal M}^{++}_{\pm\pm} = - {\cal M}^{--}_{\pm\pm}\,,
\ee
giving
\be
  \rho^{\pm\pm,\pm\pm}_{gg\to (t\bar t)_1} =  - \rho^{\pm\pm,\mp\mp}_{gg\to (t\bar t)_1}\,,
\ee
just like when toponium production is modelled as the production of a pseudo-scalar state~\cite{Fuks:2021xje}. The correlated decay distributions are determined by convoluting the production density matrix $\rho^{\sigma\bar\sigma,\sigma'\bar\sigma'}_{gg\to (t\bar t)_1}$ with the top and antitop decay density matrices $\mathrm{d}\rho^{t\to b \bar\ell^+ \nu_\ell}_{\sigma,\sigma'}$ and $\mathrm{d}\rho^{\bar t\to \bar{b} \ell^{\prime-} \bar{\nu}_\ell'}_{\bar\sigma',\bar\sigma'}$,
\be
  \sum\limits_{\sigma,\bar\sigma\phantom{'}} \sum\limits_{\sigma',\bar\sigma'}\rho^{\sigma\bar\sigma,\sigma'\bar\sigma'}_{gg\to (t\bar t)_1} \, \mathrm{d}\rho^{t\to b \ell^+ \nu_\ell}_{\sigma,\sigma'}\,  \mathrm{d}\rho^{\bar t\to \bar{b} \ell^{\prime-} \bar{\nu}_\ell'}_{\bar\sigma',\bar\sigma'}\,.
\ee
This yields
\be 
  \frac{\mathrm{d}\Gamma}{\mathrm{d}\!\cos\!\theta\, \mathrm{d}\varphi\, \mathrm{d}\!\cos\!\bar\theta\, \mathrm{d}\bar\varphi} = \frac{1}{8\pi^2} F(\bar\theta, \bar\varphi; \theta, \varphi)\,,
\ee
with
\be\label{eq:2ddistr}\begin{split}
   & F(\bar\theta, \bar\varphi; \theta, \varphi) = \frac12 \sin\bar{\theta} \sin\theta \cos(\bar{\varphi}-\varphi)\\ &\qquad +  \frac{1+\cos\bar{\theta}}{2} \, \frac{1+\cos\theta}{2} + \frac{1-\cos\bar{\theta}}{2}\, 
    \frac{1-\cos\theta}{2}\,.
\end{split}\ee
Here, $\bar{\theta}$ and $\bar{\varphi}$ ($\theta$ and $\varphi$) are the polar and azimuthal angle of the charged lepton $\ell^+$ ($\ell^{\prime-}$) in the top (antitop) rest frame, defined from the common polar $z$-axis along the top momentum direction and the common $y$-axis along the $\vec{p}_g\times \vec{p}_t$ direction, both directions being evaluated in the $gg$ (or $t\bar t$) centre-of-mass frame. The doubly-differential distribution \eqref{eq:2ddistr} has the property that
\be
  F(\theta, \varphi; \theta, \varphi) = 1 \qquad\text{and}\qquad F(\pi-\theta, \pi+\varphi; \theta, \varphi) = 0\,.
\ee
The angular distribution is thus maximum when the two leptons have the same angles in the rest frame of their parent top and antitop quarks, and minimal when they are back-to-back. Although the angles $(\bar\theta, \bar\varphi)$ and $(\theta, \varphi)$ are measured in different Lorentz frames, in the non-relativistic limit $p^*/m_t\to 0$, the two rest frames merge to the $t\bar t$ rest frame and \eqref{eq:2ddistr} reduces to
\be\label{eq:F_approx}
   F(\bar\theta, \bar\varphi; \theta, \varphi) \underset{\frac{p^*}{m_t}\to 0} \longrightarrow \frac{1+\cos\theta_{\ell^+\ell^{\prime-}}}{2}\,,
\ee
by using the trigonometric relation
\be
  \cos\theta_{\ell^+\ell^{\prime-}} = \cos\bar\theta\cos\theta + \sin\bar\theta\sin\theta \cos(\bar{\varphi}-\varphi) \,.
\ee
We therefore expect that the di-lepton angular distribution originating from toponium effects should be peaking at a small opening angle $\theta_{\ell^+\ell^{\prime-}}$.

\begin{figure*}[t]
  \centering
  \includegraphics[width=0.98\textwidth]{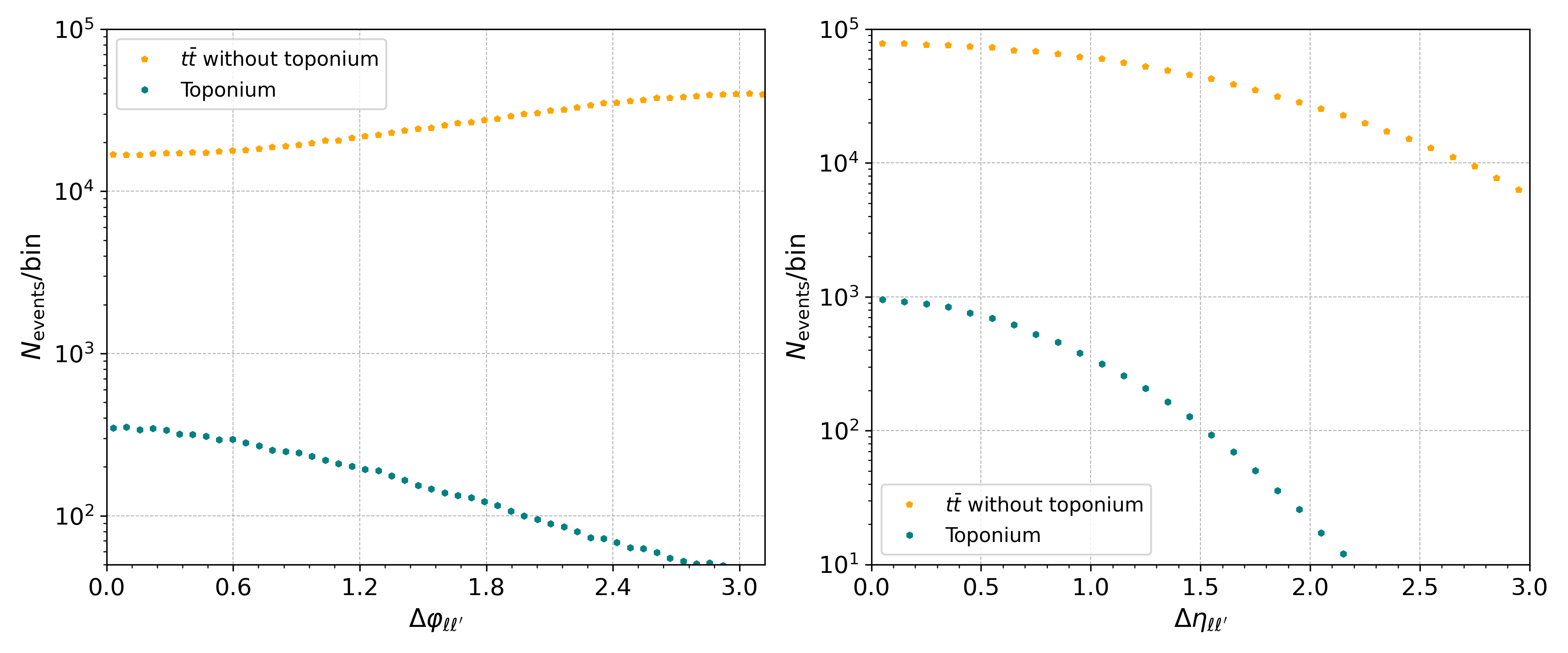}
  \caption{Distributions of the di-lepton azimuthal angle difference $\Delta\varphi_{\ell\ell'}$ (left) and pseudo-rapidity difference $\Delta\eta_{\ell\ell'}$ (right), emerging from di-leptonic $t\bar{t}$ production and decay at a centre-of-mass energy $\sqrt{s}=13$~TeV with an integrated luminosity of 140~fb$^{-1}$. Predictions are shown for conventional top-antitop production without toponium effects (orange pentagons) and for toponium production with Green's function re-weighting (teal hexagons).\label{fig:angles}}
\end{figure*}

\begin{figure}[t]
  \centering
  \includegraphics[width=0.98\columnwidth]{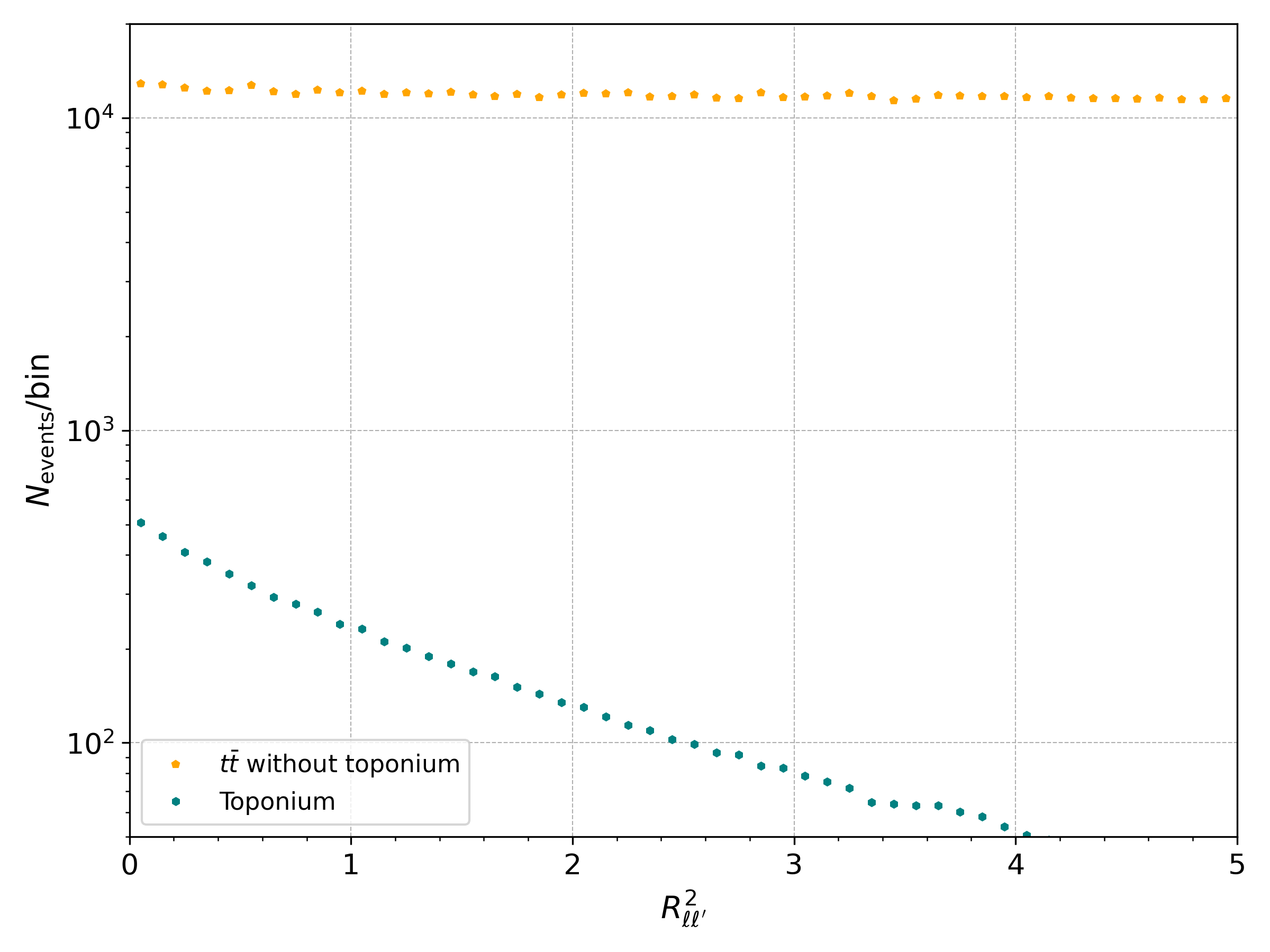}
  \caption{Same as in figure~\ref{fig:angles}, but for the di-lepton separation in the transverse plane, $R_{\ell\ell'}$.\label{fig:drll}}
\end{figure}

This is illustrated in figure~\ref{fig:angles}, in which we show the distributions of the di-lepton azimuthal angle difference $\Delta \varphi_{\ell\ell'} \equiv |\bar{\varphi}-\varphi|$ and the di-lepton pseudo-rapidity difference $\Delta \eta_{\ell\ell'} = |\eta_\ell - \eta_{\ell'}|$. We present predictions for the toponium contributions incorporating Green's function re-weighting (teal hexagons), as well as for conventional top-antitop production without toponium effects (orange pentagons). The results indicate that the toponium contribution is more pronounced in the lower bins of both the $\Delta \varphi_{\ell\ell'}$ and $\Delta \eta_{\ell\ell'}$ distributions, which is consistent with expectations from the density matrix analysis. In our earlier study~\cite{Fuks:2021xje}, we demonstrated that the $3\sigma$ discrepancy between ATLAS data and simulations~\cite{ATLAS:2019hau} was compatible with toponium formation at the LHC, by means of an analysis using the di-lepton invariant-mass ($m_{\ell\ell'}$) spectrum in addition to the $\Delta \varphi_{\ell\ell'}$ distribution. For instance, we could impose cuts such as
\be\label{eq:cut1}
  \Delta \varphi_{\ell\ell'} < 0.9\qquad\text{and}\qquad  \Delta \eta_{\ell\ell'}<0.9\,.
\ee
This strategy would yield a number of conventional top-antitop events $N_{t\bar t}$ (without toponium contributions) and a number of toponium events $N_\mathrm{toponium}$ events given by
\be
  N_{t\bar t} = 127,000 \qquad\text{and}\qquad
  N_\mathrm{toponium} = 3,520\,,
\ee
for an integrated luminosity of 140~fb$^{-1}$. This leads to a toponium signal significance $\sigma$ and signal-to-noise ratio $S/N$,
\be\begin{split}
  \sigma =&\ \frac{N_\mathrm{toponium}}{\sqrt{N_{t\bar t}+N_\mathrm{toponium}}} = 9.73 \,,\\[.2cm]
  S/N =&\ \frac{N_\mathrm{toponium}}{N_{t\bar t}} = 2.77\% \,.
\end{split}\ee
In comparison, the cuts proposed in~\cite{Fuks:2021xje},
\be\label{eq:cut2}
  \Delta \varphi_{\ell\ell'} < \frac{\pi}{5} \qquad\text{and}\qquad m_{\ell\ell'}<40~\mathrm{GeV }\,,
\ee
would give the alternative event counts $\tilde N_{t\bar t}$ and $\tilde N_\mathrm{toponium}$,
\be
  \tilde N_{t\bar t} = 77,100 \qquad\text{and}\qquad
  \tilde N_\mathrm{toponium} = 2,200\,.
\ee
This would imply a significantly smaller sensitivity of $\sigma=7.83$, for a slightly enhanced signal-to-noise ratio of $S/N=2.86\%$. 

Alternatively, the angular distributions shown in figure~\ref{fig:angles} and the approximation~\eqref{eq:F_approx} suggest that enforcing a cut on the angular separation variable $R_{\ell\ell'}^2 = (\Delta \eta_{\ell\ell'})^2 + (\Delta \varphi_{\ell\ell'})^2$ would enhance the signal relative to the conventional top-antitop background. From the corresponding distribution shown in figure~\ref{fig:drll}, we implement, instead of the cuts \eqref{eq:cut1} or \eqref{eq:cut2}, the selection
\be
  R_{\ell\ell'}^2 < 0.8\,.
\ee
This single cut on $R_{\ell\ell'}^2$ selects $\bar N_{t\bar t}$ and $\bar N_\mathrm{toponium}$ conventional top-antitop and toponium events,
\be
  \bar N_{t\bar t} = 99,400 \qquad\text{and}\qquad
  \bar N_\mathrm{toponium} = 2,980\,,
\ee
which leads to
\be
  \sigma = 9.32\qquad\text{and}\qquad S/N = 3.00\% \,.
\ee
While the significance is slightly lower than for the cuts~\eqref{eq:cut1} (but still much larger than what could be expected from our previous analysis~\cite{Fuks:2021xje}), the signal-to-noise ratio is now more favourable and reaches 3\%.

Toponium characterisation can be further achieved from a kinematic reconstruction of the system made of the two leptons, $b$-jets and missing transverse momentum, as detailed in our previous study~\cite{Fuks:2021xje}. This would require pairing each lepton $\ell_{1,2}$ with a $b$-jet. Here, $\ell_1$ and $\ell_2$ refer to the hardest and softest leptons, respectively, while we denote by $b_i$ the $b$-jet paired with lepton $\ell_i$. The correct pairing is determined by requiring that $m(\ell_1, b_1) > m(\ell_1, b_2)$. Assuming that the transverse momenta of the reconstructed top and antitop are approximately equal, which should hold for the signal, we can proceed to reconstruct the four-momenta of the two neutrinos through a kinematic fit. The resulting second-order equation for each neutrino introduces a two-fold ambiguity, which we resolve by choosing the solution that yields a reconstructed top mass closest to the known $m_t$ value. Examining distributions of observables involving the two reconstructed top quarks could then potentially be used as handles to characterise the excess events observed by the ATLAS and CMS collaborations~\cite{ATLAS:2019hau, ATLAS:2023gsl, CMS:2024ybg} as arising from toponium. We refer to \cite{Fuks:2021xje, Aguilar-Saavedra:2024mnm} for more details.

\section{Conclusion and outlook}\label{sec:conclusion}
In this paper, we revisited the impact of toponium formation on top-antitop production at the LHC, presenting a new strategy to model these effects at LO+PS. Our approach is centred on re-weighting the relevant matrix elements by ratios of QCD Green's functions, derived from solving the Lippmann-Schwinger equation for the QCD potential. This technique was implemented within the \mgamc event generator, allowing hence for an efficient incorporation of toponium effects through conventional Monte Carlo simulations typical of LHC analyses.

While including toponium effects is crucial for improving the precision of Standard Model predictions, especially given the recent high-statistics measurements at the LHC that exhibit anomalies potentially linked to the omission of the toponium contributions in the associated simulations, further work is still required. Future efforts should focus on quantifying the uncertainties inherent in our approach, extending the formalism to include higher-spin and higher-partial-wave toponium states, and matching our predictions to calculations at higher perturbative orders. Taken together, these steps will pave the way for more refined predictions for top-antitop predictions at the LHC, and allow for a deeper understanding of the non-perturbative QCD dynamics at play in top-antitop production near the threshold.

\begin{acknowledgements}
We thank Yuki Sumino for fruitful discussions on kinematic distributions of toponium-induced final states, and Thomas Waluda-Duquesne for useful comments on our Green's function tables. The work of BF was supported in part by Grant ANR-21-CE31-0013 (Project DMwithLLPatLHC) from the French \emph{Agence Nationale de la Recherche} (ANR), the one of KM by a Natural Science Basic Research Program of Shaanxi Province, China (2023-JC-YB-041), and the one of YJZ by JSPS KAKENHI Grant No.~21H01077 and 23K03403.
\end{acknowledgements}

\bibliographystyle{JHEP}
\bibliography{toponium}
\end{document}